\documentclass[%
 aip,
 amsmath,amssymb,
 reprint,%
author-year,%
]{revtex4-1}

\usepackage{graphicx}  
\usepackage{dcolumn}   
\usepackage{verbatim}   
\usepackage{xcolor}   
\usepackage{amsmath}
\usepackage{amssymb}

\usepackage{braket}
\usepackage{pstricks}
\usepackage{epsfig}

\newcommand{\Ree}{{\rm Re}}
\newcommand{\Ha}{{\rm Ha}}
\newcommand{\Rm}{{\rm Rm}}
\newcommand{\Pm}{{\rm Pm}}
\newcommand{\er}{{\bf\hat e_r}}
\newcommand{\et}{{\bf\hat e_\theta}}

\newcommand{\ez}{{\bf\hat e_z}}
\newcommand{\ve}{{\mathbf{v}}}
\newcommand{\bv}{{\bf B}}
\newcommand{\bvo}{{\bf B_0}}

\newcommand{\rv}{\ensuremath{\mathbf{r}}}
\newcommand{\lp}{\ensuremath{\left(}}
\newcommand{\rp}{\ensuremath{\right)}}

\begin{document}

\title{High dimensional tori and chaotic and intermittent transients
  in magnetohydrodynamic Couette flows}

\author{F. Garcia}
\affiliation{Department of Fluid Mechanics, Universitat Polit\`ecnica
  de Catalunya, Avda. Eduard Maristany 16, 08019 Barcelona, Spain }

\author{J. Ogbonna}
\affiliation{Department of Magnetohydrodynamics, Helmholtz-Zentrum
  Dresden-Rossendorf, Bautzner Landstra\ss e 400, D-01328 Dresden,
  Germany}

\author{A. Giesecke}
\affiliation{Department of Magnetohydrodynamics, Helmholtz-Zentrum
  Dresden-Rossendorf, Bautzner Landstra\ss e 400, D-01328 Dresden,
  Germany}

\author{F. Stefani}
\affiliation{Department of Magnetohydrodynamics, Helmholtz-Zentrum
  Dresden-Rossendorf, Bautzner Landstra\ss e 400, D-01328 Dresden,
  Germany}

\date{\today}

\begin{abstract}
The magnetised spherical Couette (MSC) problem, a three dimensional
magnetohydrodynamic paradigmatic model in geo- and astrophysics, is
considered to investigate bifurcations to high-dimensional invariant
tori and chaotic flows in large scale dissipative dynamical systems
with symmetry. The main goal of the present study is to elucidate the
origin of chaotic transients and intermittent behaviour from two
different sequences of Hopf bifurcations involving invariant tori with
four fundamental frequencies, which may be resonant. Numerical
evidence of the existence of a crisis event destroying chaotic
attractors and giving rise to the chaotic transients is provided. It
is also shown that unstable invariant tori take part in the time
evolution of these chaotic transients. For one sequence of
bifurcations, the study demonstrates that chaotic transients display
on-off intermittent behaviour. A possible explanatory mechanism is
discussed.
\end{abstract}

\maketitle

\section{Introduction}

Numerical simulations are nowadays a fundamental tool to advance in
the field of geophysical and astrophysical fluid dynamics. Canonical
models describing the motion of electrically-conducting fluids are
derived from the well-known Navier-Stokes and induction equations
(\cite{MoDo19}). These are time-dependent and nonlinear partial
differential equations whose analytic solution is unknown and
therefore numerical algorithms are developed for the approximation of
their solution with computer simulations. The analysis of simulation
data when contrasted with observations helps to develop more accurate
theoretical dynamo models of geo- and astrophysical phenomena
(\cite{RoGl00}). In addition, simulations also provide valuable
information for the design and guiding of experiments in the field
(\cite{GLPGS02}).

One of the classical examples of a geophysical and astrophysical
problem is the magnetised spherical Couette (MSC) problem in which a
conducting fluid fills the gap between two differentially rotating
spheres in the presence of a magnetic field applied parallel to the
axis of rotation. This problem has been widely studied both
numerically (\cite{Hol09,TEO11,GJG11,GSGS20}) as well as
experimentally (\cite{SMTHDHAL04,KKSS17,KNS18,BTHW18,OGGSS20,OGGSS22})
due to its relevance to understanding the origin and evolution of
planetary and stellar flows.

In the mathematical formulation of the MSC problem, three
dimensionless parameters describe the input physics. They are the
aspect ratio of the spherical shell $\chi=r_i/r_o$ ($r_i$ and $r_o$
are the inner and outer radii, respectively), the Reynolds number
$\Ree$ measuring the differential rotation, and the Hartmann number
$\Ha$ measuring the strength of the applied magnetic field. As these
parameters are varied, a rich variety of dynamical regimes have been
discovered, described, and analysed during the last decades thanks to
the use of numerical simulations
(e.\,g. \cite{HoSk01,Hol09,TEO11,GJG11,Kap14,KNS18,GSGS20}). For
instance, for a given relatively thin shell geometry ($\chi\ge 0.5$)
and for all magnetic field strengths ($\Ha$), a time independent and
axisymmetric (i.\,e. invariant by any azimuthal rotation) base state
is found for sufficiently small differential rotation ($\Ree$). With
increasing $\Ree$, this base state becomes unstable to a
non-axisymmetric instability, with some azimuthal symmetry $m=m_1$
($m$ is the azimuthal wave number), whose spatial nature strongly
depends on $\Ha$. For low $\Ha$, the instability is basically
hydrodynamic and situated at the radial jet flowing outwards in the
equatorial plane. At moderate $\Ha$, the instability takes the form of
a return flow concentrated in the middle of the shell, and for larger
$\Ha$, the flow streamlines are basically parallel to the axis of
rotation, giving rise to the shear-layer instability.

These dynamical regimes are subjected to the symmetries of the MSC
system, which is an {\bf{SO}}$(2)\times${\bf{Z}}$_2$ equivariant
system, meaning that it is invariant to azimuthal rotations and
reflections with respect to the equatorial plane. Theoretical studies
in the framework of dynamical systems with symmetry
(\cite{CrKn91,EZK92,GoSt03}) have demonstrated that the Hopf-like
instability of the axisymmetric base state gives rise to rotating
waves (RW), a type of periodic flow with certain azimuthal symmetry
$m=m_1$ for which the time evolution is described as a solid body
rotation of a fixed flow pattern (\cite{GaSt18}). A secondary Hopf
bifurcation is responsible for the appearance of modulated rotating
waves (MRW) which are quasiperiodic flows with two fundamental
frequencies, i.\,e. invariant two dimensional tori
(\cite{Ran82,GLM00,CaJo12,GSGS19}). Successive Hopf bifurcations, for
which a symmetry breaking occurs, give rise to high-dimensional
invariant tori with three (\cite{GSGS20}) and even four fundamental
frequencies (\cite{GSGS20b}). The analysis of symmetry breaking
bifurcations of MRW in symmetric systems is of fundamental importance
since they enable the Newhouse-Ruelle-Takens theorem (\cite{NRT78}) to
be overcome in the route to chaotic flows.

This paper concentrates on the analysis of high-dimensional tori (with
dimensions of 2,3 and 4) and the bifurcations to chaotic flows, which
have been recently discovered by \cite{GSGS20b}, but also investigates
the presence of transient phenomena associated with the disappearance
of a chaotic attractor. An exploration of the parameter space leads us
to the identification of resonant invariant four-dimensional tori and
also to detect and investigate intermittent behaviour during chaotic
transients. Invariant tori, and their bifurcations to chaotic flows,
are widely studied in the case of low dimensional systems. For
instance, the appearance and disappearance of four-dimensional tori
and chaotic attractors have been studied in detail by \cite{KST11} for
a system of three coupled van der Pol oscillators. Quasiperiodic
invariant tori, including resonant motions and global bifurcation to
chaotic flows, have also been comprehensively investigated by
\cite{FlJa20} in the case of a three dimensional dissipative vector
field. For high-dimensional systems arising from the discretisation of
partial differential equations the computational methods of regular
states require the implementation of parallel algorithms based on
advanced numerical linear algebra methods
(e.\,g. \cite{DWCDDEGHLSPSSTT14,Tuc20}). For this reason the
exploration of both the parameter as well as the phase space in this
type of problems is restricted by the computational resources and thus
there exist few studies addressing global bifurcations of
high-dimensional tori and chaotic flows. This is in particular the
case for the MSC system considered in our study.

A crisis event (\cite{GOY82,GOY83}) is one of the most common
mechanisms by which a chaotic attractor loses stability in favour of a
chaotic saddle. The latter comprises nonattracting invariant sets in
the phase space which are responsible for chaotic transient
phenomena. Transient chaos is however not only associated with chaotic
saddles but may be also due to the existence of a blow-out bifurcation
(\cite{OtSo94}), and even other mechanisms (see~\cite{OmTe22} for a
review). The global properties of chaotic attractors, the structure of
their basin of attraction (\cite{SoOt93}), and the bifurcations giving
rise to them in nonlinear systems, are nowadays an active area of
research. This is especially true in the case of low-dimensional
systems such as the Duffing-Van der Pol oscillator and related
problems (e.\,g. \cite{YXW13,Fen17}), but also for problems in other
disciplines such as economics (\cite{LoNu02}).

Chaotic motion can be intermittent, in the sense that the time
evolution of the system alternates between two or more different
states that occupy distinct regions in the phase space (see the review
of \cite{KnMo99}). Different types of intermittent dynamics, type-I,
II, and III, were studied in \cite{PoMa80} in the context of the
Lorenz problem with the aim of shedding light on the observed
behaviour in fluid dynamics experiments. In their description, the
different types of intermittence arise from the different ways in
which a simple fixed point of the system loses stability as a control
parameter is varied. Intermittent behaviour may also arise as a result
of certain types of crisis for which a chaotic attractor suddenly
changes its distribution in the phase space (\cite{GORY87}). Unstable
invariant objects, lying on an invariant manifold of the phase state,
organise on-off intermittent behaviour defined in \cite{PST93}. In
this latter study, the fact that the stability of these unstable
orbits is controlled by dynamics outside from the manifold resulted in
a key issue. The special type of on-off intermittency may appear as a
result of a blow-out bifurcation (\cite{OtSo94,OSAV95}) when the
chaotic attractor losing stability does not have a riddled basin of
attraction. The existence of intermittent behaviour has also been
associated with an orbit riddling (\cite{He05}), after a tangency in the
phase space is developed from high-dimensional tori, once the
parameter is varied in a one dimensional nonlinear wave system.

In the fluid dynamics context, such as the plane Couette or pipe flows
(\cite{EFSS08,BDH19,Let17}) the analysis of transient chaos and
chaotic saddles has provided valuable insight in the understanding of
the transition to turbulence, the formation of coherent structures,
and the characterisation of intermittent behaviour. In the concrete
case of shear flows, the concept of intermittency has been of
importance to understand transition to turbulence
(\cite{AvHo13,Lem_et_al16}). Depending on the aspect ratio of the
container the transition to turbulence could have a spatio-temporal
nature, or in contrast, could be mainly described as a purely temporal
process (\cite{PhMa11}).  The existence of these unstable states and
their invariant manifolds (e.\,g. \cite{GHC08,KUL12}) is key for
describing the laminar-turbulant boundary and coherent structures in
shear flows. According to \cite{VeKa11}, a homoclinic tangle between
the stable and unstable manifolds is responsible for chaotic
intermittent bursting. Later, \cite{ChPa13} conjectured that the
existence of bursts can be also described by heteroclinic connections
approaching an unstable periodic orbit. A systematic computation and
description of heteroclinic connections, a structurally stable
solution departing from unstable equilibria or periodic orbits and
tending to another unstable regular flow, has been performed in
\cite{HGCV09}.

The analysis of intermittent chaotic motion in three dimensional MHD
systems in the framework of dynamical systems theory is also
fundamental for understanding how cosmic magnetic fields are generated
(\cite{PBM13}) but also to shed light into the formation of coherent
structures in astrophysical plasmas (\cite{MWSGOOD22}). In the case of
numerical simulations within a periodic box
(e.\,g. \cite{SOALF01,AlPo08}), dynamo action takes place as a result
of a blow-out (non-hysteretic) bifurcation with associated on-off
intermittency. This is also the case for the MSC dynamo problem
(\cite{RaDo13}), in which the magnetic field is not externally applied
as in our case. However, in the case of the $\alpha^2$ dynamo model of
\cite{ORCK21}, intermittent behaviour was not found and the dynamo
effect saturates thanks to a hysteretic blow-out bifurcation.  Dynamo
experiments in a turbulent background have also shown intermittent
magnetic field measurements (\cite{NSKJF06}).

This paper starts by describing the MSC system of equations and the
numerical methods employed to solve them in Sec.~\ref{sec:eq_num}. The
selected output data and the methods employed to analyse it are
outlined in Sec.~\ref{sec:ana}.  The paper then mainly addresses two
important aspects described above. Firstly, we provide further
evidence of the bifurcation scenario first described in~\cite{GSGS20b}
by presenting a new branch, which follows a similar sequence of
symmetry breaking bifurcations, giving rise to invariant tori with
four fundamental frequencies, including resonant motions. These
solutions are analysed in detail and their structure in the phase
space is characterised in Sec.~\ref{sec:bif_diag}. Secondly, the
investigation of chaotic attractors and the appearance of chaotic
transients is performed in Sec.~\ref{sec:int_trans}. In this section,
we provide numerical evidence of the existence of a crisis event
destroying chaotic attractors and giving rise to chaotic
transients. For a certain range of Hartmann numbers, the transients
involve on-off intermittent motion, which may be due to a tangency in
the phase space. Section~\ref{sec:int_stat} contains the verification
of on-off signature of these chaotic transients using standard
statistical methods. Finally, Sec.~\ref{sec:con} summarises the
results obtained.

\section{Numerical model}
\label{sec:eq_num}

\begin{figure}
\begin{center}
\hspace{0.mm}\includegraphics[width=0.95\linewidth]{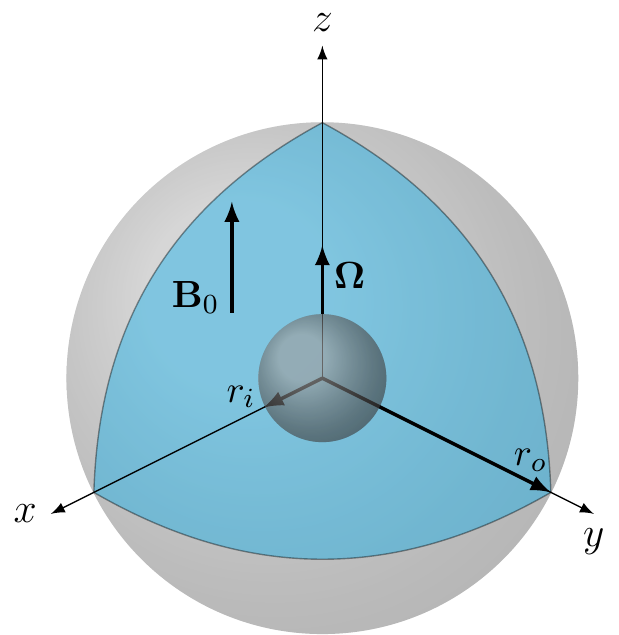}
\end{center}  
\caption{Sketch of the spherical shell including the applied magnetic
  field and the inner sphere rotation.}
\label{fig:sph}
\end{figure}

The fluid container is a spherical shell of inner and outer radii
$r_i$ and $r_o$, respectively. The fluid is conducting with constant
physical properties: electrical conductivity $\sigma=1/(\eta\mu_0)$
($\mu_0$ is the free-space magnetic permeability), density $\rho$,
kinematic viscosity $\nu$, and magnetic diffusivity $\eta$. The outer
sphere is kept fixed whereas the inner sphere rotates around a
vertical axis with constant angular velocity $\Omega$ (see figure
\ref{fig:sph}). The whole system is exposed to a uniform axial
magnetic field $\bvo=B_0 \cos(\theta)\er-B_0 \sin(\theta)\et$ of
amplitude $B_0$ ($\theta$ is the colatitude).

To obtain a dimensionless system of equations, the characteristic
quantities $d=r_o-r_i$, $d^2/\nu$, $r_i\Omega$ and $B_0$ are employed
as scales for length, time, velocity and magnetic field,
respectively. For moderate rotation rates of the inner sphere
(moderate Reynolds $\Ree=\Omega r_i d/\nu\sim 10^3$), the
inductionless approximation can be used for liquid metals, for
instance, the eutectic alloy GaInSn, since the magnetic Prandtl number
$\Pm=\nu/\eta\sim O(10^{-6})$ (\cite{MBCN08}) and then the requirement
for a low magnetic Reynolds number $\Rm=\Omega r_i d/\eta=\Pm\Ree \sim
10^{-3}\ll1$ is satisfied.

In the inductionless approximation the magnetic field is decomposed as
$\bv=\ez+\Rm{\bf b}$ and the terms $O(\Rm)$ are neglected in the
Navier-Stokes and induction equations
\begin{align*}
&\partial_t\ve+\Ree\lp\ve\cdot\nabla\rp\ve =
-\nabla p+\nabla^2\ve+\Ha^2(\nabla\times {\bf b})\times\ez,\\
&\nabla\times(\ve\times\ez)+\nabla^2{\bf b}=0,\\
&\nabla\cdot\ve=0, \quad \nabla\cdot{\bf b}=0.
\end{align*}
This system of equations is commonly referred to as the MSC system and
depends on three parameters: the Reynolds number, the Hartmann number,
and the aspect ratio
\begin{equation*}
  \Ree=\frac{\Omega r_i d}{\nu}, \quad
  \Ha=\frac{B_0d}{\sqrt{\mu_0\rho\nu\eta}}=B_0d\sqrt{\frac{\sigma}{\rho\nu}},\quad
  \chi=\frac{r_i}{r_o}.
\end{equation*}
The flow boundary conditions are no-slip ($v_r=v_\theta=v_\varphi=0$)
at the outher sphere ($r=r_o$) and constant rotation
($v_r=v_\theta=0,~v_\varphi= \sin{\theta}$) at the inner sphere
($r=r_i$). Insulating exterior regions are selected for the magnetic
field. This is the usual setting in previous numerical studies
(e.\,g. \cite{HoSk01,GSGS20}) and experiments
(e.\,g. \cite{KKSS17,OGGSS20}).

The numerical method used to solve the MSC system is outlined next. A
detailed description for discretisation and time stepping of the
Navier-Stokes equations in rotating spherical geometry can be found
in~\cite{GNGS10} and references therein. The velocity $\ve$ and
magnetic ${\bf b}$ divergence-free fields are expressed in terms of
toroidal and poloidal potentials (\cite{Cha81}). The expression for
the velocity field is
\begin{equation}
  \ve=\nabla\times\lp\Psi\rv\rp+\nabla\times\nabla\times\lp\Phi\rv\rp,
\label{eq:pot}  
\end{equation}
$\rv=r~\er$ being the position vector. The scalar potentials are
expanded as spherical harmonics series (up to degree $L_{\text{max}}$
and order $M_{\text{max}}=L_{\text{max}}$) in the angular coordinates
$(\theta,\varphi)$ ($\varphi$ is the longitude).  In the case of the
velocity potentials the spherical harmonics expansion is
\begin{eqnarray}
  \Psi(t,r,\theta,\varphi)=\sum_{l=0}^{L_{\text{max}}}\sum_{m=-l}^{l}{\Psi_{l}^{m}(r,t)Y_l^{m}(\theta,\varphi)},\label{eq:serie_psi}\\
  \Phi(t,r,\theta,\varphi)=\sum_{l=0}^{L_{\text{max}}}\sum_{m=-l}^{l}{\Phi_{l}^{m}(r,t)Y_l^{m}(\theta,\varphi)},\label{eq:serie_phi}
\end{eqnarray}
with $Y_l^{m}(\theta,\varphi)=P_l^m(\cos\theta) e^{im\varphi}$,
$P_l^m$ being the normalised associated Legendre functions, of degree
$l$ and order $m$, satisfying $\Psi_l^{-m}=\overline{\Psi_l^{m}}$. The
potentials are unique if the condition $\Psi_0^0=\Phi_0^0=0$ is
imposed.  A collocation method, on a Gauss--Lobatto mesh of $N_r$
points, is employed for the discretisation of the radial
direction. OpenMP parallel strategies and optimised libraries (FFTW3,
see \cite{FrJo05}) and matrix-matrix products (dgemm, see
\cite{GoGe08}) are implemented in the numerical code. The time
integration method, based on high order implicit-explicit backward
differentiation formulas (IMEX--BDF), has an explicit treatment of the
Lorenz force and the nonlinear terms.

\section{Analysis of MHD flows}
\label{sec:ana}

The interpretation of MHD flows is based on the analysis of very long
time series obtained with direct numerical simulations (DNS) of the
MSC system. The different MHD flows are computed by varying the
magnetic field amplitude and keeping the rotation of the inner sphere
and the aspect ratio of the shell fixed. This translates in only varying
$\Ha$ and keeping the values of $\Ree$ and $\eta$ fixed at $\Ree=10^3$
and $\eta=0.5$, respectively.

The numerical resolutions employed for this study are $n_r=40$ radial
collocation points and a spherical harmonic truncation parameter of
$L_{\text{max}}=84$. This resolution is tested by increasing the
values to $n_r=60$ and $L_{\text{max}}=126$ to verify that a solution
with the same temporal dependence, and with similar (within 1\% of
error) time-averaged properties, is obtained. Most DNS are evolved for
more than 100 viscous time units to filter initial transients before
the attractor is reached.

The procedure to obtain the branches of solutions is standard. A
saturated solution (i.\,e. a solution after the initial transient) at
$\Ha_1$ is employed as initial condition for simulating the solution
at $\Ha_2=\Ha_1+\delta \Ha$ ($\delta \Ha$ may be positive or
negative). The first solution is a modulated rotating wave (MRW)
obtained from a rotating wave (RW) with azimuthal symmetry $m=4$
already computed in~\cite{GaSt18}.

We use an azimuthally constrained DNS code to compute unstable flows
with certain azimuthal symmetry $m=m_1$. The only requirement is that
the unstable manifold of the flow lies away from the subset $m=m_1$,
i.\,e. the unstable mode cannot have azimuthal symmetry $m=km_1$ for
any integer $k$. We recall that the flow has azimuthal symmetry
$m=m_1$ if and only if the spherical harmonics expansion of the scalar
fields (see Eq.~\ref{eq:serie_psi}) has only non-vanishing amplitudes
on the azimuthal wave numbers $m=km_1$, for any integer $k$.

The analysis of the solutions is performed from the time series
extracted from the DNS. Concretely, we analyse the volume-averaged
kinetic energy
\begin{equation}
K=\frac{1}{2{\mathcal V}}\int_{\mathcal V} \ve
\cdot \ve \;dv,
\label{eq:ener_dens}
\end{equation}
where $\mathcal V$ is the shell volume and $\ve$ is the velocity
field. We also consider the kinetic energies $K_m$ associated with
each azimuthal wave number $m$, obtained by only considering the
spherical harmonics amplitudes $\Psi^m_l$ and $\Phi^m_l$ of a single
order $m=m_0$ and degree $l$ satisfying $|m_0|\le l \le
L_{\text{max}}$ and setting to zero all the other amplitudes with
$m\ne m_0$. The analysis of $K_m$ for different azimuthal modes $m$
provides information about how energy is distributed in space and
helps to identify the most energetic wave number $m_{\text{max}}$
which satisfies $\overline{K}_{m_{\text{max}}}> \overline{K}_m$, $1\le
m\le L_{\text{max}},~\text{and}~m\ne m_{\text{max}}$, with the
overline representing a time average.  The non-axisymmetric kinetic
energy $K_{\text{na}}$ is computed considering only the $m\ne 0$ wave
numbers in the spherical harmonics expansion of the potential fields
and measures the departure of the solutions from the purely
axisymmetric ($m=0$) base flow. The poloidal kinetic energy
$K^{\text{P}}$ or the non-axisymmetric toroidal kinetic energy
$K^{\text{T}}_{\text{na}}$ are defined by only using the poloidal
scalar or the $m\ne 0$ azimuthal wave numbers of the toroidal scalar,
respectively, in the expression of the velocity field. While the
poloidal scalar is directly related to the radial component of the
velocity, the toroidal scalar only contributes to the azimuthal and
colatitudinal components.

As discussed in the introductory section, the system is
{\bf{SO}}$(2)\times${\bf{Z}}$_2$-equivariant (invariant to azimuthal
rotations and reflections with respect to the equatorial plane). These
systems are associated with a particular type of solutions called
rotating waves (RW) and modulated rotating waves (MRW), which can be
precisely defined in terms of their spatio-temporal symmetry
(e.\,g. \cite{Ran82,GLM00,SGN13,Bud_etal17,GSGS19}).

A solution of the system $u(t,r,\theta,\varphi)$ is a rotating wave
(RW) if $u(t,r,\theta,\varphi)=\mathcal{R}(\omega
t)u(0,r,\theta,\varphi)$, where $\mathcal{R}(\omega t)$ is a rigid
rotation about the vertical axis of angle $\omega t$. Then, the RW is
described as a solid body rotation at a constant angular velocity
$\omega$ of a given pattern $u(0,r,\theta,\varphi)$. Although a RW is
periodic, its azimuthally-averaged properties are constant. If the
solution is a modulated rotating wave,
$u(t,r,\theta,\varphi)=\mathcal{R}(\omega
t+\gamma(t))\tilde{u}(t,r,\theta,\varphi)$, where $\gamma$ and
$\tilde{u}$ are $\tau$-periodic functions of time. In this case, the
temporal dependence of the solution is described by two fundamental
frequencies $\omega$ and $2\pi/\tau$. Because of this particular
spatio-temporal symmetry, the azimuthally-averaged properties are
periodic.

Rotating waves develop once the base axisymmetric state becomes
unstable to non-axisymmetric perturbations. Secondary Hopf-type
bifurcations of RW give rise to MRW with two fundamental frequencies
(invariant tori). Successive Hopf-type bifurcations result in MRW with
three and even four fundamental frequencies as studied
in~\cite{GSGS20b}. To identify quasiperiodic MRW with $N_f$
fundamental frequencies and azimuthal symmetry $m=m_1$, the notation
$N_f$T-MRW$_{m_1}$ is used. These quasiperiodic flows are investigated
by means of a Fourier based refined analysis (\cite{Las93}) and
Poincar\'e sections as will be subsequently outlined.

As commented above, the main frequency of RW and MRW, corresponding to
the azimuthal drift of the wave, is removed from the frequency
spectrum of volume-averaged quantities. These quantities are constant
in time for RW, periodic for 2T-MRW, quasiperiodic with two
fundamental frequencies for 3T-MRW, and quasiperiodic with three
fundamental frequencies for 4T-MRW. For this reason the Poincar{\'e}
sections of volume-averaged quantities for 2T-MRW correspond to a
single point, to a closed curve for 3T-MRW , and to a surface for
4T-MRW. Poincar\'e sections of weakly chaotic flows (consisting of a
cloud of points) may resemble surface-like sections of regular
4T-MRW. To fully distinguish both types of solutions, we have
performed the chaos test of~\cite{LFC92,Las93b} and studied the
diffusion measure of the solution in the phase space. In addition, we
have checked that the three frequencies $f_1$, $f_2$, and $f_3$ are
fundamental, in the sense that any other frequency $f_j$ of the
spectrum of any volume-averaged kinetic energy is a linear combination
of the fundamental frequencies with integer coefficients.


\section{Bifurcation diagrams of MRW}
\label{sec:bif_diag}

\begin{figure}
\begin{center}
\hspace{0.mm}\includegraphics[width=0.95\linewidth]{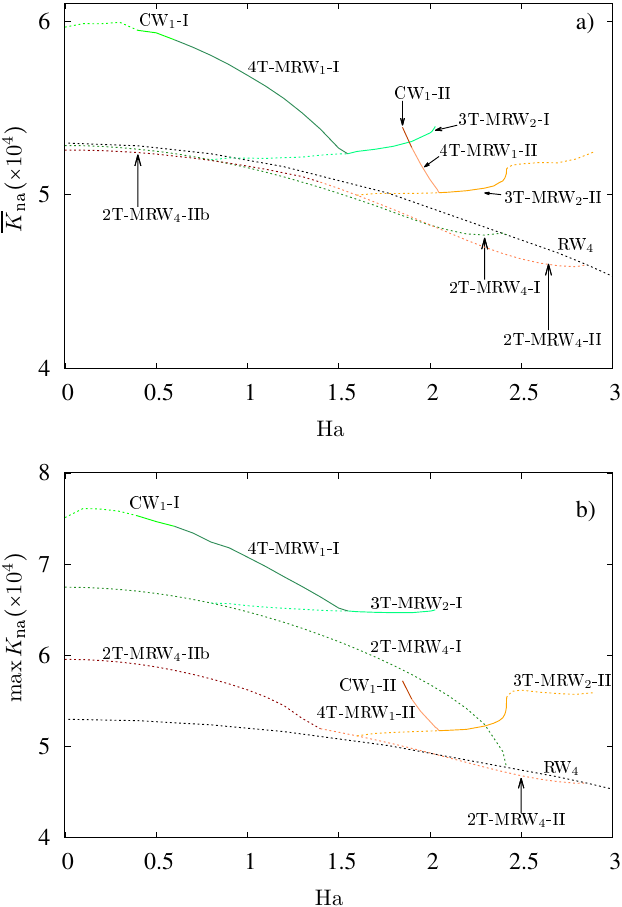}
\end{center}  
\caption{Bifurcation diagram of the volume-averaged non-axisymmetric
  kinetic energy density $K_{\text{na}}$ versus $\Ha$. (a)
  Time-averaged $K_{\text{na}}$ and (b) maximum values of
  $K_{\text{na}}$.  Solid (dashed) lines are used for stable
  (unstable) flows. Branches of rotating waves RW$_m$, modulated
  rotating waves MRW$_m$, and chaotic waves CW$_m$, the subscript $m$
  being their azimuthal symmetry, are displayed. MRWs with $N_f$
  fundamental frequencies are labelled as $N_f$T, with
  $N_f=2,...,4$. The colours distinguish the types of solutions.}
\label{fig:bif_diag}
\end{figure}

The time-averaged and maximum values of the non-axisymmetric kinetic
energy of saturated flows at $\eta=0.5$ and $\Ree=10^3$ are plotted in
figure~\ref{fig:bif_diag} versus $\Ha$. In this figure, there are two
different branches of 2T-MRW which bifurcate, without breaking the
$m=4$ symmetry, from the branch of RW with azimuthal symmetry $m=4$
(originally computed by~\cite{GaSt18}). These branches are labelled as
2T-MRW$_4$-I and 2T-MRW$_4$-II. Notice that for these types of
solutions, the time average is very similar to that of the parent
RW$_4$ (Fig.~\ref{fig:bif_diag}(a)), but as they have oscillatory
(periodic) $K_{\text{na}}$, the maximum value is clearly different
(see Fig.~\ref{fig:bif_diag}(b)).

As described by~\cite{GSGS20b}, for the branch I there is a sequence
of Hopf bifurcations giving rise to MRW with four fundamental
frequencies (branch 4T-MRW$_1$):
$$\text{RW}_4
\xrightarrow[]{\raisebox{.5pt}{\textcircled{\raisebox{-.9pt} {1}}}}
\text{2T-MRW}_4
\xrightarrow[]{\raisebox{.5pt}{\textcircled{\raisebox{-.9pt} {2}}}}
\text{3T-MRW}_2
\xrightarrow[]{\raisebox{.5pt}{\textcircled{\raisebox{-.9pt} {3}}}}
\text{4T-MRW}_1,$$\\[2.mm] which is exactly the same sequence obtained
in figure~\ref{fig:bif_diag} for the branch II. The critical Hartmann
numbers for the bifurcations on branch I are:
$$\Ha_1\approx 2.4, \quad \Ha_2\approx 0.8, \quad \Ha_3\approx 1.52,$$
and on branch II:
$$\Ha_1\approx 2.86, \quad \Ha_2\approx 1.6, \quad \Ha_3\approx
2.05.$$ Both sequences (I and II) of bifurcations involve the same
sequence of azimuthal symmetry breaking and appearance of stable solutions:
\begin{align*}
&m=4&\xrightarrow[]{\raisebox{.5pt}{\textcircled{\raisebox{-.9pt} {1}}}}& ~~~m=4&\xrightarrow[]{\raisebox{.5pt}{\textcircled{\raisebox{-.9pt} {2}}}}& ~~~m=2 &\xrightarrow[]{\raisebox{.5pt}{\textcircled{\raisebox{-.9pt} {3}}}}& ~~~m=1&\\
&\text{Unst.}&\xrightarrow[]{\raisebox{.5pt}{\textcircled{\raisebox{-.9pt} {1}}}}& ~~~\text{Unst.} &\xrightarrow[]{\raisebox{.5pt}{\textcircled{\raisebox{-.9pt} {2}}}} & ~~~\text{Unst.} & \xrightarrow[]{\raisebox{.5pt}{\textcircled{\raisebox{-.9pt} {3}}}} & ~~~\text{St.}&
\end{align*}  
Bifurcation $\raisebox{.5pt}{\textcircled{\raisebox{-.9pt} {1}}}$ is
supercritical whereas bifurcations
$\raisebox{.5pt}{\textcircled{\raisebox{-.9pt} {2}}}$ and
$\raisebox{.5pt}{\textcircled{\raisebox{-.9pt} {3}}}$ are
subcritical. Although for both branches the same sequence of
bifurcations occurs, their solutions have different structures in the
phase space. This is shown in the next two sections, where each branch
is described in detail.

\begin{figure*}
\hspace{0.mm}\includegraphics[width=0.95\linewidth]{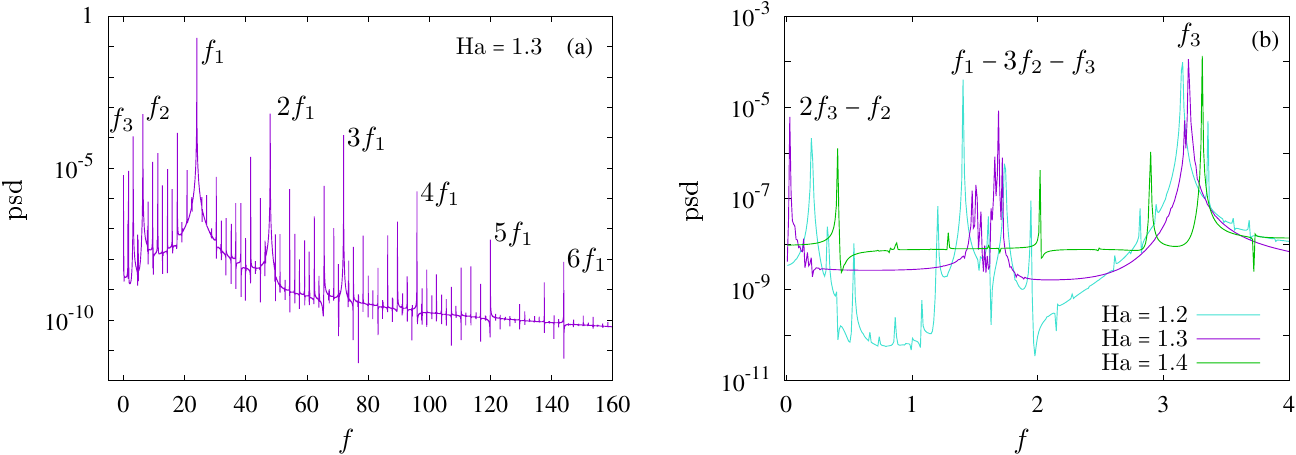}
\caption{{\bf Frequencies on branch 4T-MRW$_1$-I:} (a) Power spectral
  density (psd) for the volume-averaged kinetic energy density $K_4$
  of the $m=4$ mode for the 4T-MRW$_1$ at $\Ha=1.3$. (b) Detail of
  (a) including the psd of the 4T-MRW$_1$ at $\Ha=1.2$ and at
  $\Ha=1.4$.}
\label{fig:4T_freq}      
\end{figure*}

\subsection{Branch I}
\label{sec:brI}

This branch was first studied in~\cite{GSGS20b}, but the transition to
chaotic flows (labelled as CW$_1$) was not clarified due to the time
series, although very long (up to $100$ time units), not being
sufficiently long to cope with very small frequencies exhibited by
some solutions along the branch of 4T-MRW$_1$-I (around $\Ha=1.35$) as
we show in the following.  The power spectral density (psd) of $K_4$
is displayed in figure~\ref{fig:4T_freq}(a) for a 4T-MRW$_1$-I at
$\Ha=1.3$ and a detail of the lower frequencies including the psd of
4T-MRW$_1$-I at $\Ha=1.2$ and at $\Ha=1.4$ are displayed in
figure~\ref{fig:4T_freq}(b).  The minimum frequencies are
$f_{\text{min}}=0.205$ for $\Ha=1.2$, $f_{\text{min}}=0.0311$ for
$\Ha=1.3$, and $f_{\text{min}}=0.412$ for $\Ha=1.4$, i.\,.e, the value
at $\Ha=1.3$ is almost an order of magnitude smaller than those at the
neighbouring $\Ha$.We have checked that the smallest frequencies of
the psd correspond to $f_{\text{min}}=2f_3-f_2$, where $f_1$, $f_2$,
and $f_3$ are the fundamental frequencies labelled in
fig.~\ref{fig:4T_freq}(a).

\begin{figure}
\hspace{0.mm}\includegraphics[width=0.9\linewidth]{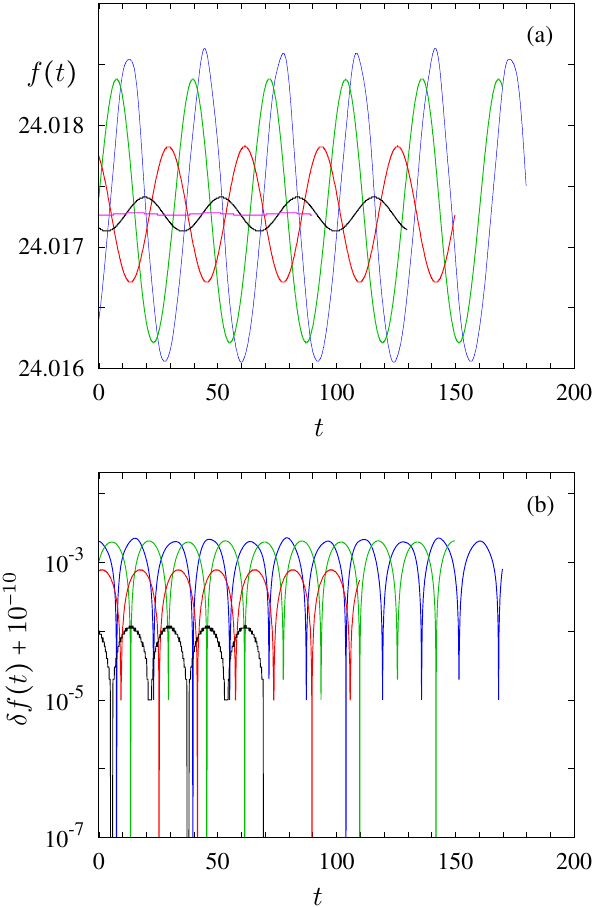}
\caption{{\bf Diffusion measure on branch 4T-MRW$_1$-I:}
  Time-dependent frequency spectrum based on Laskar algorithm
  (SDDSToolKit). The time series corresponds to the volume-averaged
  kinetic energy of the $m=4$ component of the flow.  (a) Frequency
  with maximum amplitude versus time. (b) Time difference $\delta
  f(t)=|f(t+T)-f(t)|$ versus time (logarithmic scale). Different
  colors denote different lengths of the time series (blue $T=10$,
  green $T=20$, red $T=40$, black $T=60$, and magenta $T=100$). The
  solution corresponds to a 4T-MRW$_1$ at $\Ha=1.3$ with a very
  small frequency peak (Fig.~\ref{fig:4T_freq}(b)).}
\label{fig:4T_lask}      
\end{figure}

To demonstrate that the solution at $\Ha=1.3$ is a regular, i.\,e. a
4T-MRW$_1$ and not a chaotic flow, the diffusion measure of the orbit
in the phase space is estimated from a time dependent accurate
frequency analysis following the algorithm of~\cite{LFC92,Las93b},
which has been successfully applied to the MSC problem
in~\cite{GSGS21}. The procedure requires the selection of a time
window $T<T_f$ ($T_f$ is the total time interval of the time series)
and the computation of the first fundamental frequency $f_1$ using
Laskar's algorithm~\cite{Las93}, which achieves a relative accuracy of
around $10^{-5}$, over the time window $T$. The solution is considered
regular if the variation of $f_1$ over the different time windows
(covering the total time interval) is smaller than the considered
accuracy. Figure~\ref{fig:4T_lask}(a) displays the value of $f_1$
versus time for different time windows $T=10,20,40,60,100,$ and a final
time of $T_f=200$. This figure clearly shows that only for $T=60$ and $T=100$
the variation of the frequency is insignificant. This variation is
quantified in Fig~\ref{fig:4T_lask}(b) by employing the time
difference $\delta f(t)=|f(t+T)-f(t)|$ which only can be computed for
$t<T_f-2T$ (see~\cite{GSGS21} for details). For $T=60$, values of
$\delta f(t)\approx 10^{-4}$ are obtained whereas for $T\le 40$ (used
in~\cite{GSGS20b}) the time difference rises to $\delta f(t)\approx
10^{-3}$. Since only time windows $T\le 40$ over a total time interval
of $T_f=100$ were used in~\cite{GSGS20b}, the solution was not
classified as regular. With time series spanning up to $T_f=200$, we
have been able to confirm the regular character of all the solutions
on branch 4T-MRW$_1$ down to $\Ha=0.7$. With the same length of the
time series, the solution obtained at $\Ha=0.6$ gives rise to
variations of order $10^{-2}$ for frequencies computed with time
windows of length $T=100$.  This ensures that the solution is chaotic
so the transition to chaotic flows CW$_1$-I may be approximated to
occur at $\Ha\approx 0.65$.

Because $f_1$, $f_2$ and $f_3$ are fundamental, any other frequency
$f_j$ is a linear combination of $f_1$, $f_2$ and $f_3$ with integer
coefficients. The relative error value for the linear combinations of
the frequencies $f_j$ is
$\epsilon_j=|f_{j}-k^1_jf_{1}-k^2_jf_{2}-k^3_jf_{3}|/f_{j}$. A linear
combination is valid whenever $\epsilon_j<5\times 10^{-5}$ in
accordance with the relative accuracy ($10^{-5}$) achieved for the
frequencies (see~\cite{GSGS21} for a detailed study of this
accuracy). We observe that as $\Ha$ is decreased along the branch the
number of $f_j$ with larger amplitudes increases so that larger
integers $k_i$ are required for the linear combinations. We have used
$6\le k_i\le 18$ and considered the frequencies $f_j$ with amplitudes
larger than $10^{-6}A_{\text{max}}$, where $A_{\text{max}}$ is the
largest amplitude of the psd.

\begin{figure}
\hspace{0.mm}\includegraphics[width=0.9\linewidth]{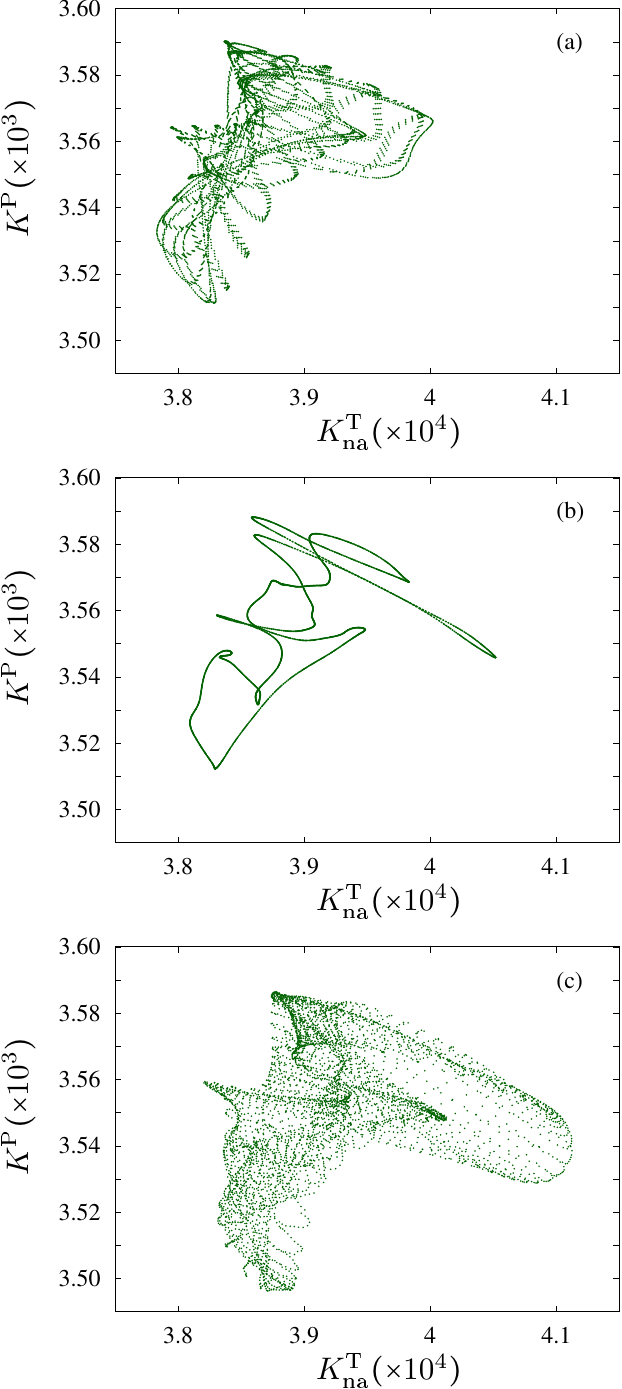}
\caption{{\bf Resonances on branch 4T-MRW$_1$-I:} Poincar\'e sections
  at the time instants $t_i$ defined by the constraint
  $K(t_i)=0.02177$, $K$ being the volume-averaged kinetic energy. The
  volume-averaged poloidal kinetic energy $K^{\text{P}}(t_i)$ is
  displayed versus the volume-averaged toroidal non-axisymmetric energy
  $K^{\text{T}}_{\text{na}}(t_i)$. The solutions are 4T-MRW$_1$ at
  (a) $\Ha=0.85$, (b) $\Ha=0.8$, and (c) $\Ha=0.75$. }
\label{fig:4T_poinc}      
\end{figure}

As noticed above for $\Ha=1.3$, there may be linear combinations of
the fundamental frequencies which become very small, and even vanish,
as $\Ha$ is decreased along the branch 4T-MRW$_1$. This is because the
fundamental frequencies vary with $\Ha$ and so do the linear
combinations. When a linear combination having at least one
coefficient equal to unity vanishes, a resonant solution is obtained
with one less fundamental frequency. This is the case for $\Ha=0.8$
where the relative error of the resonance is
$\epsilon_{\text{res}}=(f_{1}-4f_{2}+f_{3})/f_{1}=8.3\times
10^{-8}$. For this solution the Poincar\'e section involving
volume-averaged kinetic energies will be a closed curve. This is shown
in figure~\ref{fig:4T_poinc} for the Poincar\'e section at the time
instants $t_i$ where $K(t_i)=0.02177$ (recall that $K$ is the
volume-averaged kinetic energy). The volume-averaged poloidal kinetic
energy $K^{\text{P}}(t_i)$ is displayed versus the volume-averaged
non-axisymmetric toroidal energy $K^{\text{T}}_{\text{na}}(t_i)$ for
$\Ha=0.85$ (a), $\Ha=0.8$ (b), and $\Ha=0.75$ (c). At $\Ha=0.8$ the
Poincar\'e section is a complicated but closed curve evidencing the
resonance condition.

\subsection{Branch II}
\label{sec:brII}

As commented on before, this branch follows the same sequence of
bifurcations as branch I, but an additional branch of 2T-MRW$_4$
(labelled IIb) develops from 2T-MRW$_4$-II by means of a period
doubling bifurcation at $\Ha_{\text{pd}}\approx 1.4$ (see
figure~\ref{fig:bif_diag}). Period doubling bifurcations of 2T-MRW are
common in the MSC system at low $\Ha$. They have been described
by~\cite{GSGS20} for flows with azimuthal symmetry $m=2$ and $m=3$.

At the bifurcation point
$\raisebox{.5pt}{\textcircled{\raisebox{-.9pt} {3}}}$, the solutions
on the branch 3T-MRW$_2$-I become stable. The same happens for branch
II. By increasing $\Ha$ from this point along the branch of
3T-MRW$_2$-I becomes unstable at around $\Ha\gtrsim 2$. For $\Ha=2.05$
or $\Ha=2.1$ and by taking as the initial condition the solution at
$\Ha=2$ on the 3T-MRW$_2$-I branch, a solution on the 3T-MRW$_2$-II is
obtained after an initial transient when integrating the MSC equations
with constrained azimuthal symmetry $m=2$. As the branches
3T-MRW$_2$-I and 3T-MRW$_2$-II have $m=2$ azimuthal symmetry, the
eigenfunction at the bifurcation point (on branch I) should have at
least $m=2$ azimuthal symmetry. This gives evidence for a connection
between the unstable and stable manifolds of branch 3T-MRW$_2$-I and
branch 3T-MRW$_2$-II, respectively. This behaviour is not found for
branch II, where unstable solutions of 3T-MRW$_2$-II, arising due to a
local bifurcation at $\Ha_{\text{sn}}\approx 2.42$, can be integrated.

\begin{figure*}
  \hspace{0.mm}\includegraphics[width=0.43\linewidth]{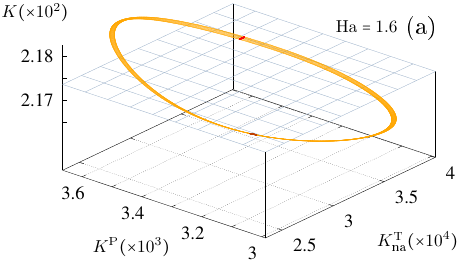}
  \hspace{0.mm}\includegraphics[width=0.43\linewidth]{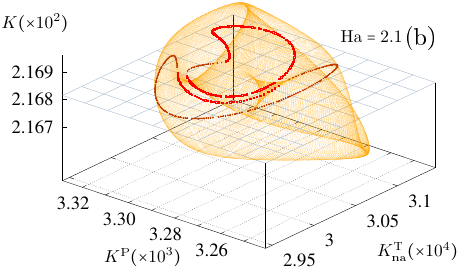}\\[5.mm]
  \hspace{0.mm}\includegraphics[width=0.43\linewidth]{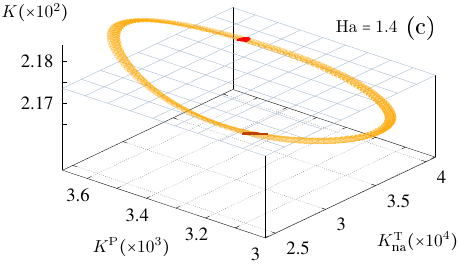}
  \hspace{0.mm}\includegraphics[width=0.43\linewidth]{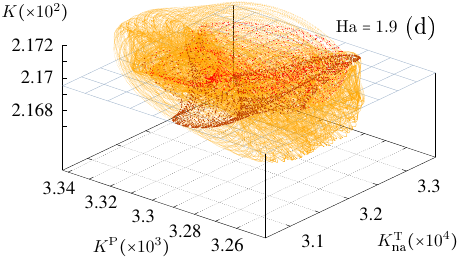}
  \caption{{\bf Phase space structure of branches I and II:} Phase space
    portraits (orange online) and Poincar\'e sections (brown and red
    online) at the time instants $t_i$ defined by the constraint
    $K(t_i)=\overline{K}$, $K$ being the volume-averaged kinetic
    energy and $\overline{K}$ its time average. The volume-averaged
    poloidal kinetic energy $K^{\text{P}}(t_i)$ is displayed versus
    the volume-averaged toroidal non-axisymmetric energy
    $K^{\text{T}}_{\text{na}}(t_i)$ in the horizontal coordinates. (a)
    Solution on the 3T-MRW$_2$-I branch at $\Ha=1.6$. (b) Solution on
    the 3T-MRW$_2$-II branch at $\Ha=2.1$. (c) Solution on the
    4T-MRW$_1$-I branch at $\Ha=1.4$. (d) Solution on the
    4T-MRW$_1$-II branch at $\Ha=1.9$.}
\label{fig:3d_poinc}      
\end{figure*}

Despite the solutions on branches I and II following the same sequence of
bifurcations and originating from the same branch of RW$_4$, the orbits of
the solutions have significantly different distributions in the phase
space, revealing different degree of complexity. Notice also that the
existence of 4T-MRW$_1$-II is restricted to a small $\Ha$
interval. The three-dimensional plots of figure~\ref{fig:3d_poinc}
correspond to the phase portraits of volume-averaged kinetic energies
(see figure caption for details) including a Poincar\'e section (the
grid plane) for a 3T-MRW$_2$ and a 4T-MRW$_1$ solutions on branch I
(panels (a) and (c)) and on branch II (panels (b) and (d)). While the
solutions on branch I have a phase portrait with a clear toroidal-like
structure, the phase portraits of solutions on branch II are more
folded and intricate with spherical-like shape.


In the framework of a sequence of bifurcations giving rise to 3T and
4T (high-dimensional tori) in the one dimensional nonlinear system of
wave propagation of~\cite{He05}, the appearance of bursting
intermittent solutions was associated with a tangency of the orbit in
the phase space that allows an occasional riddling, giving rise to the
intermittent behaviour. In our case this tangency has already happened
on branch 3T-MRW$_1$-II at around $\Ha\approx 2.25$. This is clearly
seen on figure~\ref{fig:3d_poinc} (b) for a 3T-MRW$_1$-II at
$\Ha\approx 2.1$ where the two patches of the Poincar\'e section
overlap as in \cite{He05}.

As for branch I, the transition to chaotic flows is investigated by
means of a time-dependent refined frequency analysis requiring long
time integrations over $100$ time units. We have estimated the
critical Hartmann number for this transition to be $\Ha\approx
1.88$. Chaotic flows, either on branches I or II, are stable over a
relatively small interval of $\Ha$. For branch I they lose stability
at $\Ha\approx 0.45$ and for branch II at $\Ha\approx 1.75$. By
further decreasing $\Ha$ on each branch, chaotic flows become unstable
giving rise to large initial transients. The sudden disappearance of a
chaotic attractor as the parameter of the system is varied usually
corresponds to a boundary crisis (\cite{GOY82}), which occurs as the
chaotic attractor collide with unstable fixed points, periodic orbits,
or invariant tori (~\cite{GOY83}).  By taking a look at the
bifurcation diagram of branch I (fig.~\ref{fig:bif_diag}), it may be
that the collision of CW$_1$-I branch occurs with the unstable
4T-MRW$_1$-I branch, born when the branch 4T-MRW$_1$-I loses
stability. This could be possible since both branches may have similar
values of $\overline{K_{\text{na}}}$ close to the crisis. A similar
situation seems to happen with branch II.

\begin{figure}
  \hspace{0.mm}\includegraphics[width=1.\linewidth]{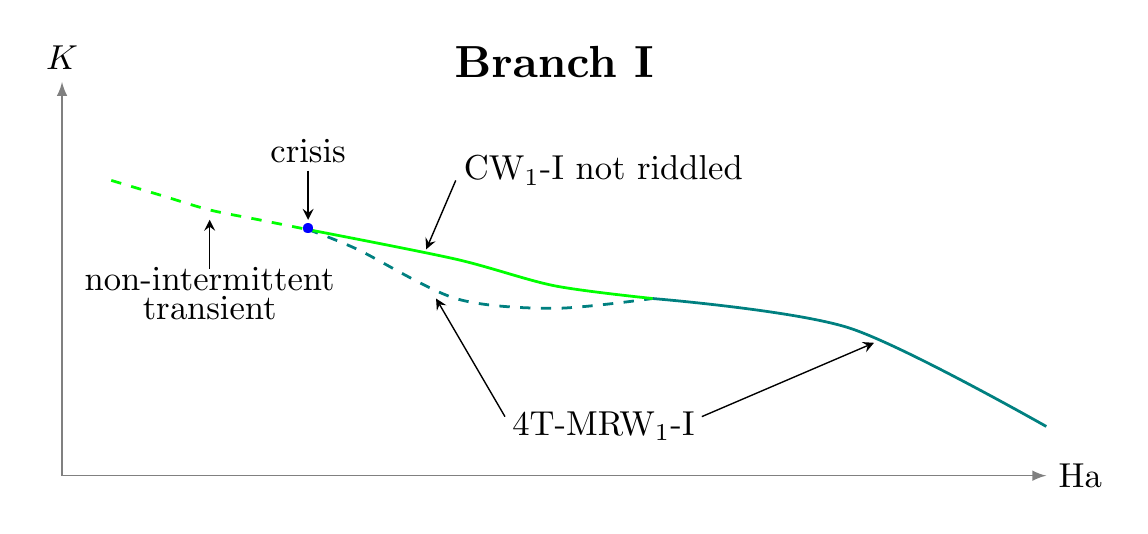}\\[5.mm]
  \hspace{0.mm}\includegraphics[width=1.\linewidth]{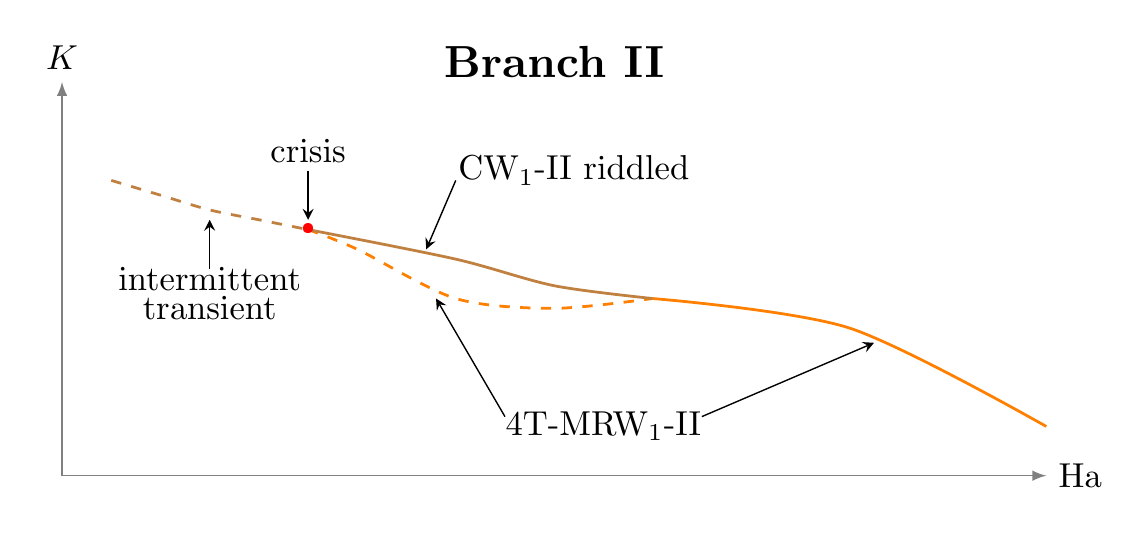}
  \caption{{\bf Possible crisis scenario for branches I and II:} Schemes
    of the bifurcation diagrams displaying a crisis event of the
    chaotic waves in branches I (top) and II (bottom). The difference
    between branch I and branch II is that chaotic waves for branch II are
    riddled and that the transients after the crisis display
    intermittent behaviour.}
\label{fig:bif_sch}      
\end{figure}

We remark that chaotic flows may also lose stability by means of a
blow-out bifurcation (\cite{OtSo94}), for which the chaotic state lies
within an invariant manifold. This will be the case, for instance,
when the chaotic attractor has certain spatial symmetries
(\cite{OSAV95}), which does not correspond to our situation (chaotic
flows have lost all azimuthal symmetries). In addition, blow-out
bifurcations may as well give rise to chaotic transients when the
chaotic attractor before the bifurcation has a riddled basin of
attraction. Conversely, if the basin of the chaotic attractor losing
stability is not riddled, the blow-out bifurcation gives rise to
intermittent behaviour (\cite{OtSo94}). In our case, the situation is
just the reverse. For branch II, the basin of attraction of the
chaotic attractor seems to be riddled (due the tangency in the phase
space), but as it is shown in the next section, we observe
intermittent behaviour. For branch I, the basin of attraction is not
riddled but intermittent phenomena seems not to be present. For all
these reasons, we tend to believe that chaotic flows lose stability by
means of a crisis and not by means of a blow-out
bifurcation. Schematic diagrams summarising the possible scenarios for
branches I and II are shown in Fig~\ref{fig:bif_sch}.

\section{Transient behaviour}
\label{sec:int_trans}

This section is devoted to the description and analysis of initial
transients, which can span long time intervals. They are sometimes
found when taking initial conditions on the branches of chaotic waves
CW$_1$-I and II. As commented in the previous section, the existence
of these long initial transients seems to be related with the
occurrence of a boundary crisis at which a chaotic attractor loses
stability developing a chaotic saddle (\cite{GOY82,GOY83}). The time
evolution of initial conditions lying close to these chaotic saddles,
which are invariant sets repelling in some directions of the phase
space (\cite{KaGr85}), is characterised by a long-lived chaotic
behaviour that suddenly stops and reaches a stable, usually regular
attractor, afterwards. Recently, the existence of a terminal transient
phase just before the appearance of the final attractor was made
evident in~\cite{LiPa18}.

\begin{figure*}
  \hspace{0.mm}\includegraphics[width=0.49\linewidth]{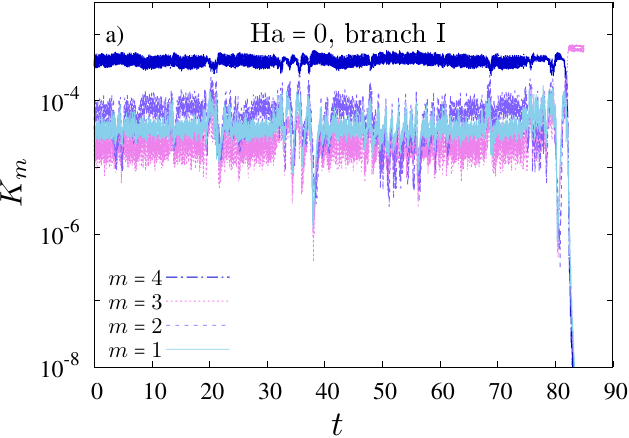}
  \hspace{0.mm}\includegraphics[width=0.49\linewidth]{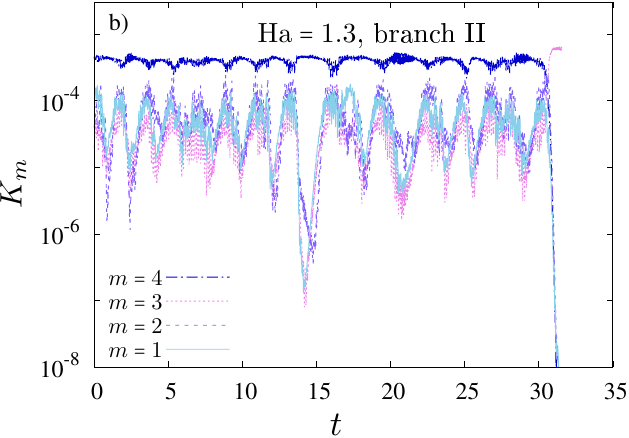}  
\caption{{\bf Chaotic transients from branches I and II:} Kinetic
  energy $K_{m}$ of the azimuthal wave numbers $m=1,2,3,4$ versus
  time. (a) $\Ha=0$ corresponding to branch I. (b) $\Ha=1.3$
  corresponding to branch II.}
\label{fig:mener}
\end{figure*}

The sudden disapearance of the long-lived chaotic transient behaviour
and the convergence towards a regular attractor is visualised in
Fig.~\ref{fig:mener}. The time evolution of the kinetic energies $K_m$
for $m=1,..,4$ during the transient phase close to chaotic saddles
corresponding to branch I (from now on called saddles I) is displayed
in Fig.~\ref{fig:mener}(a) for $\Ha=0$. The same plot is shown in
Fig.~\ref{fig:mener}(b) for the transient phase close to chaotic
saddles corresponding to branch II (saddles II) for $\Ha=1.3$.  In
both panels, the dominant azimuthal wave number is $m=4$, since
$K_4\gg K_m$ ($m=1,2,3$). On average, $K_4$ may be more than one order
of magnitude larger than $K_m$ ($m=1,2,3$), and about two orders of
magnitude larger at some intervals. For these time intervals, the
transient phase almost lies in the invariant subspace $m=4$ (i.\,e. it
is quasi-invariant). Notice that in figure~\ref{fig:mener}, the
lifetime of the transient phase (i.\,e. the limits of the horizontal
axis) varies with $\Ha$. This lifetime also strongly depends on the
initial condition considered, but statistically tends to infinity when
the critical parameter of the crisis is approached
(\cite{GORY87}). The main difference between figure~\ref{fig:mener}(a)
and figure~\ref{fig:mener}(b) is that in the latter (i.\,e. for the
transient phases close to saddles II), the time intervals where
$K_4>10K_m$ ($m=1,2,3$), i.\,e. the time intervals where the solution
lies in the quasi-invariant subspace, are longer and more
frequent. Nevertheless, both long-lived transients ultimately saturate
to a regular solution, a 2T-MRW$_3$ (with $m=3$ azimuthal symmetry)
described previously in \cite{GSGS20}, since $K_4,K_2,K_1\rightarrow
0$ in Fig.~\ref{fig:mener}.


As $\Ha$ is decreased, chaotic saddles may develop homoclinic or
heteroclinic tangencies of their invariant manifolds, which tend to
increase the fractal dimension of the saddle in a stair-case way
(\cite{LZG99}), giving rise to more complex chaotic
transients. Numerical evidence of this behaviour is provided later in
this section. Although unstable, the investigation of these transient
states is important since they may exist for very long times and thus
may be the only states that can be detected experimentally. These
transient states exhibit a temporal behaviour which can be related
with the existence of unstable MRW belonging to branches I and
II. Specifically, the transients of branch I (those occurring for
$\Ha<0.45$) seem to sometimes approach the unstable branch of
2T-MRW$_4$-I, whereas the transients corresponding to branch II (those
occurring for $0.9<\Ha<1.7$) may approach either the unstable branches
2T-MRW$_4$-IIb, 3T-MRW$_2$-I, or 3T-MRW$_2$-II.

\begin{figure}
\hspace{0.mm}\includegraphics[width=0.99\linewidth]{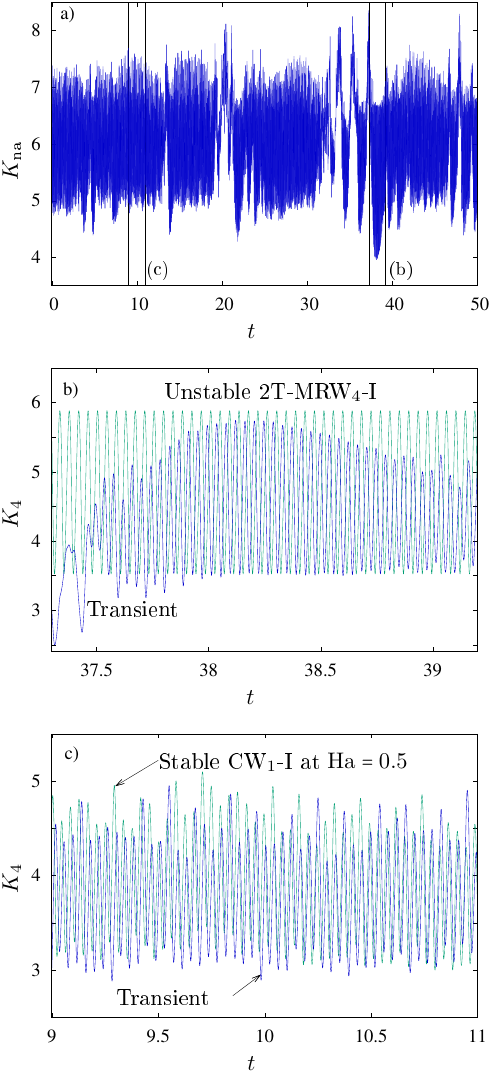}    
\caption{{\bf Chaotic transient at $\Ha=0.0$:} (a)
  Non-Axisymmetric kinetic energy , $K_{\text{na}}$ versus time for the
  transient at $\Ha=0.0$. (b) Comparison between $K_4$ of the
  transient and $K_4$ of the unstable two tori 2T-MRW$_4$-I, both at
  $\Ha=0.0$. (c) Comparison between $K_4$ of the transient at
  $\Ha=0.0$ and $K_4$ of the stable CW$_1$ at $\Ha=0.5$. The time
  intervals of (b,c) are marked in (a).}
\label{fig:Ha0.0ener}
\end{figure}

\begin{figure}
\hspace{0.mm}\includegraphics[width=0.99\linewidth]{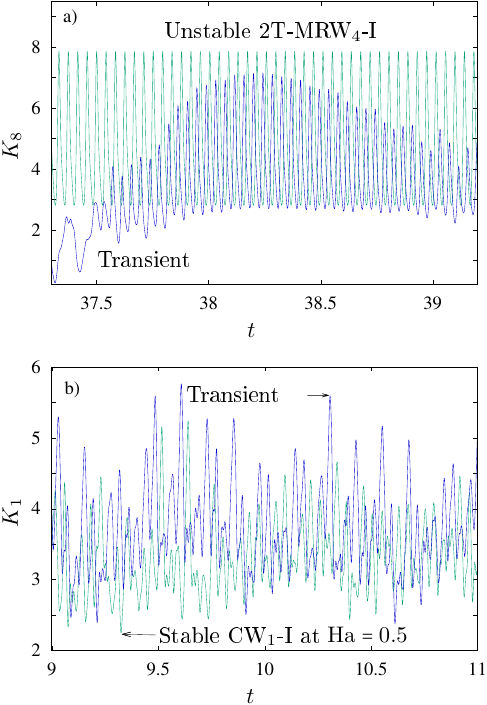}
\caption{{\bf Chaotic transient at $\Ha=0.0$:} (a) Comparison between $K_8$ of
  the transient and $K_8$ of the unstable 2T-MRW$_4$-I, both at
  $\Ha=0.0$. (c) Comparison between $K_1$ of the transient at
  $\Ha=0.0$ and $K_1$ of the unstable CRW$_1$ at $\Ha=0.5$. The time
  intervals of (a,b) correspond to those of (b,c) in
  Fig.~\ref{fig:Ha0.0ener}.}
\label{fig:Ha0.0ener_b}      
\end{figure}

\begin{figure}
  \hspace{0.mm}\includegraphics[width=0.99\linewidth]{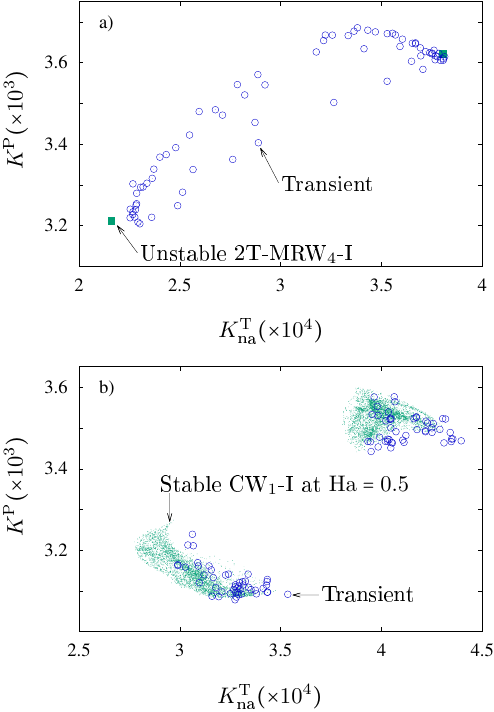}
\caption{{\bf Chaotic transient at $\Ha=0.0$:} Poincar\'e sections defined by
  the constraint $K(t)=\overline{K}$, $K$ being the volume-averaged
  kinetic energy and $\overline{K}$ its time average. The
  volume-averaged poloidal kinetic energy $K^{\text{P}}$ is displayed
  versus the volume-averaged toroidal non-axisymmetric energy
  $K^{\text{T}}_{\text{na}}$. (a) Poincar\'e sections for the
  transient (restricted to the time interval (b) of
  Fig.~\ref{fig:Ha0.0ener}, circles) and for the unstable 2T-MRW$_4$-I
  (squares), both at $\Ha=0.0$.  (b) Poincar\'e sections for the
  transient at $\Ha=0.0$ (restricted to the time interval (c) of
  Fig.~\ref{fig:Ha0.0ener}, circles) and for the stable CRW$_1$ at
  $\Ha=0.5$ (dots).  Because volume-averaged properties are
  considered, the Poincar\'e sections of 2T are a single point.}
\label{fig:Ha0.0poinc}
\end{figure}

Figure~\ref{fig:Ha0.0ener}(a) displays the non-axisymmetric kinetic
energy $K_{\text{na}}$ versus time for the transient at $\Ha=0.0$
(i.\,e. corresponding to saddle I). The evolution of $K_{\text{na}}$
seems to alternate between periods of chaotic oscillatory behaviour
(marked with the (c) label) and periods of regular oscillations
(marked with the (b) label). By comparing the kinetic energy of the
most energetic mode ($K_4$) of the transient with that of an unstable
2T-MRW$_4$-I at $\Ha=0.0$ (see Fig.~\ref{fig:Ha0.0ener}(b)) during the
time interval marked with (b) in Fig.~\ref{fig:Ha0.0ener}(a), it is
clear that the transient approaches the unstable 2T-MRW$_4$-I, since
the amplitude and period of the oscillations of $K_4$ are very
similar. The same variables are displayed in
Fig.~\ref{fig:Ha0.0ener}(c) during a chaotic period (time interval (c)
of Fig.~\ref{fig:Ha0.0ener}(a)) of the transient together with a
stable CW$_1$-I solution at $\Ha=0.5$. In this case, the transient
strongly resembles the chaotic attractor before the crisis. To provide
further evidence of the relation between the transient and the
unstable 2T-MRW$_4$-I or the chaotic attractor before the crisis, the
time series of their kinetic energies of the $m=8$ mode and that of
the $m=1$ mode are displayed in Fig.~\ref{fig:Ha0.0ener_b}(a) and
Fig.~\ref{fig:Ha0.0ener_b}(b), respectively. The agreement between the
transient time series and the other shown in the figure is as clear as
in Fig.~\ref{fig:Ha0.0ener}.

Poincar\'e sections of volume-averaged kinetic energies help to
confirm the dynamics close to the saddle. They are defined by the
constraint $K(t)=\overline{K}$, with $K$ being the volume-averaged
kinetic energy and $\overline{K}$ its time average. The
volume-averaged poloidal kinetic energy $K^{\text{P}}$ is displayed
versus the volume-averaged toroidal non-axisymmetric energy
$K^{\text{T}}_{\text{na}}$ at the Poincar\'e sections in
Fig.~\ref{fig:Ha0.0poinc}. Panel (a) contains the comparison of the
transient, during the period (b) of Fig.~\ref{fig:Ha0.0ener}, with the
unstable 2T-MRW$_4$-I (the two square points) at $\Ha=0.0$. Although
the transient points (circles) are scattered all over the plot (as
they should), they tend to be concentrated very close to the square
points corresponding to the Poincar\'e section of the
2T-MRW$_4$-I. Similarly, the pair of clouds of points (circles) of
Fig.~\ref{fig:Ha0.0poinc}(b) corresponding to the transient have
similar topologies  and they are located over the two patches of the
Poincar\'e section of the stable CW$_1$-I solution at $\Ha=0.5$.

\begin{figure}
\hspace{0.mm}\includegraphics[width=0.99\linewidth]{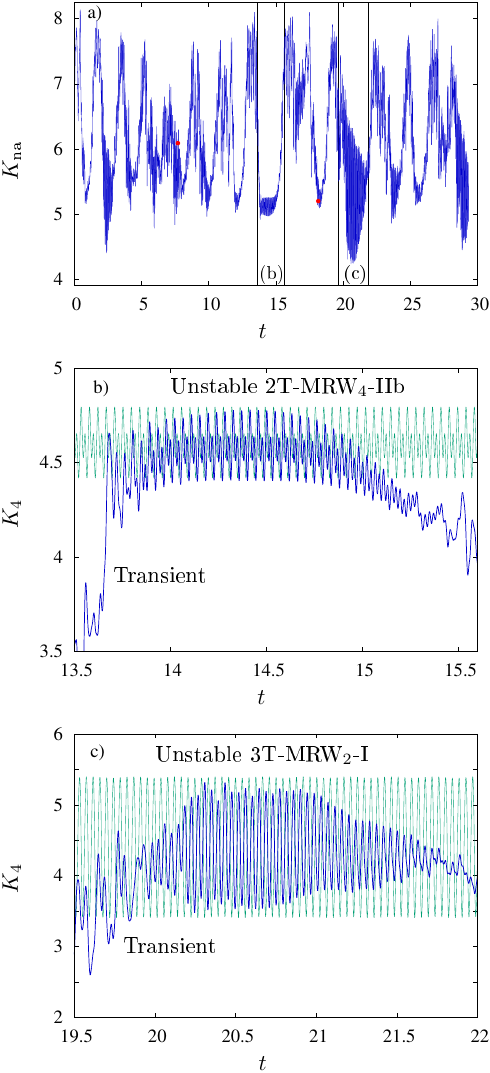}    
\caption{{\bf Chaotic transient at $\Ha=1.3$:} (a) Non-axisymmetric
  kinetic energy , $K_{\text{na}}$ versus time for the transient at
  $\Ha=1.3$. The points (red online) correspond to the snapshots of
  Fig.~\ref{fig:Ha1.3cpl} (b) Comparison of $K_4$ for the transient
  and the unstable 2T-MRW$_4$-IIb on a time interval indicated in (a).
  (c) Comparison of $K_4$ for the transient and the unstable
  3T-MRW$_2$-I on a time interval indicated in (a).}
\label{fig:Ha1.3ener}
\end{figure}

\begin{figure*}[t!]
\hspace{0.mm}\includegraphics[width=0.24\linewidth]{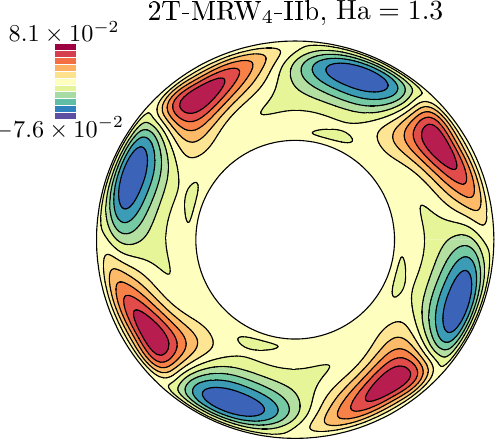}
\hspace{0.mm}\includegraphics[width=0.24\linewidth]{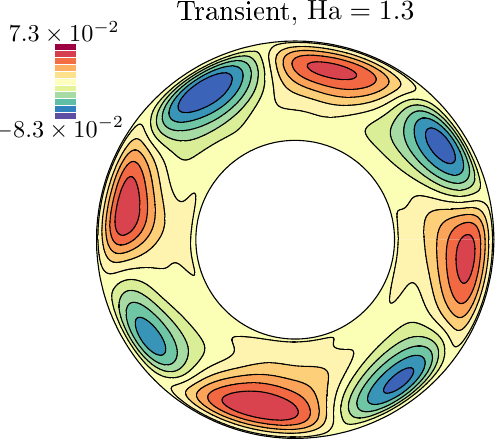}
\hspace{0.mm}\includegraphics[width=0.24\linewidth]{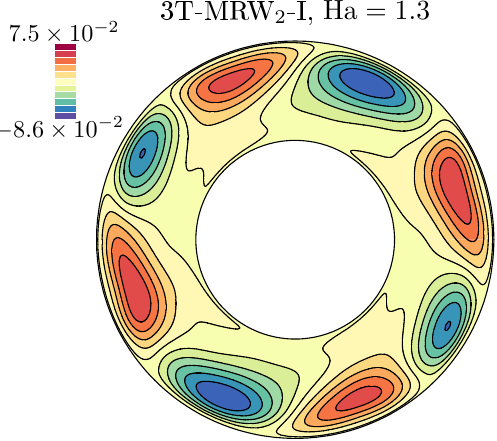}
\hspace{0.mm}\includegraphics[width=0.24\linewidth]{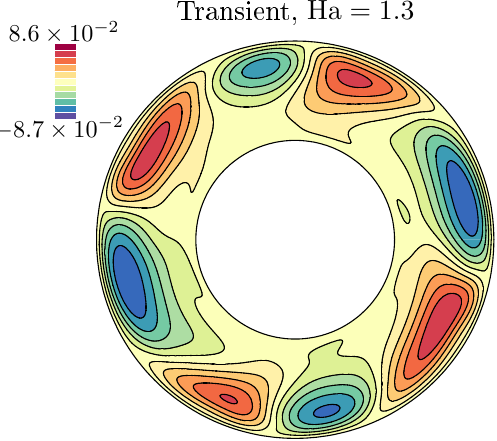}        
\caption{{\bf Chaotic transient at $\Ha=1.3$:} Snapshots showing the
  contour plots of the radial velocity on a colatitudinal slice
  slightly above the equator. From left to right the contour plots
  correspond to the unstable 2T-MRW$_4$-IIb at $\Ha=1.3$, the
  transient at $\Ha=1.3$ (point on the right of
  Fig.~\ref{fig:Ha1.3ener}(a)) with almost $m=4$ azimuthal symmetry,
  the unstable 3T-MRW$_2$-I, and the transient at $\Ha=1.3$ (point on
  the left of Fig.~\ref{fig:Ha1.3ener}(a)) with almost $m=2$ azimuthal
  symmetry, respectively.}
\label{fig:Ha1.3cpl}
\end{figure*}

\begin{figure}
\hspace{0.mm}\includegraphics[width=0.99\linewidth]{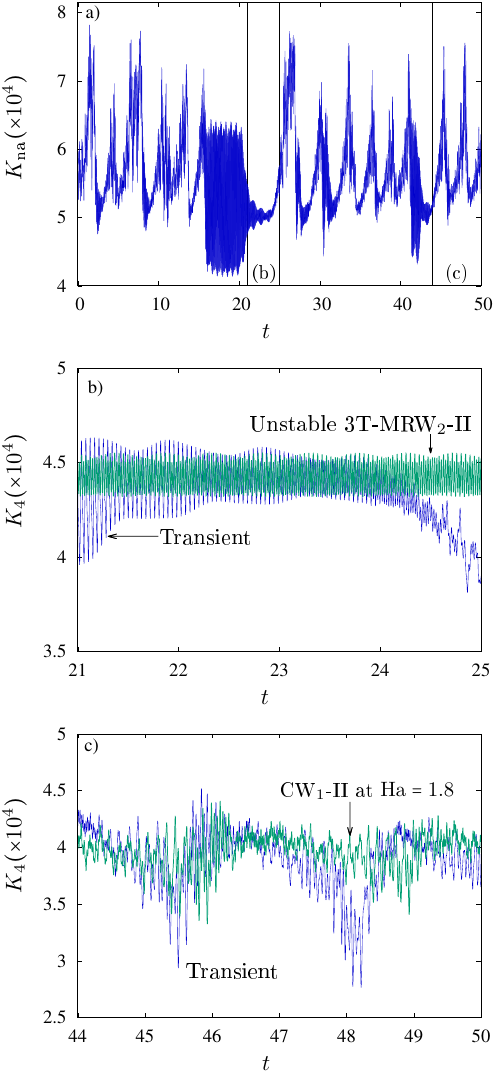}    
\caption{{\bf Chaotic transient at $\Ha=1.7$:} (a) Non-axisymmetric kinetic
  energy , $K_{\text{na}}$ versus time for the transient at
  $\Ha=1.7$. (b) Comparison of $K_4$ for the transient and the
  unstable 3T-MRW$_2$-II on a time interval indicated in (a).  (c)
  Comparison of $K_4$ for the transient and the stable CW$_1$-II at
  $\Ha=1.8$ on a time interval indicated in (a).}
\label{fig:Ha1.7ener}
\end{figure}

Figures~\ref{fig:Ha1.3ener} and \ref{fig:Ha1.7ener} are analogous to
Fig.~\ref{fig:Ha0.0ener}, but for transients at $\Ha=1.3$ and
$\Ha=1.7$, respectively, associated with the chaotic saddle II. For
$\Ha=1.3$, there are different periods of time where the solution
approaches either an unstable 2T-MRW$_4$-IIb or an unstable
3T-MRW$_2$-II (both at $\Ha=1.3$). This is shown in
Fig.~\ref{fig:Ha1.3cpl}, which compares the contour plots of the
radial velocity of the transient at some time instants with the
corresponding contour plots of the unstable 2T-MRW$_4$-IIb or the
unstable 3T-MRW$_2$-II, both at $\Ha=1.3$. In
Fig.~\ref{fig:Ha1.3ener}(b), the period doubling giving rise to the
unstable 2T-MRW$_4$-IIb branch is also reproduced during the transient
phase. For the transient at $\Ha=1.7$, besides the periods approaching
the 3T-MRW$_2$-II (fig.~\ref{fig:Ha1.7ener}(b)), there are periods
reminiscent of the chaotic attractor before the crisis
(fig.~\ref{fig:Ha1.7ener}(c)). Notice that in this case, the transient
neither can approach the branch of 2T-MRW$_4$-IIb or 3T-MRW$_2$-I
since these branches are born at $\Ha<1.6$.

Figure~\ref{fig:Ha1.7ener}(a) seems to display oscillatory regular
behaviour for approximately $t\in[15,20]$, just before the time period
marked with the label (b), i.\,e. before the transient approaches the
unstable 3T-MRW$_2$-II branch. In addition, for $t\in[15,20]$, the
transient at $\Ha=1.7$ has a symmetry of nearly $m=2$. This may be an
indication that the transient is approaching the branch 3T-MRW$_2$-I
as we argue in the following. As commented in Sec.~\ref{sec:brII}, the
unstable invariant manifold of branch 3T-MRW$_2$-I seems to be
connected with the stable invariant manifold of branch 3T-MRW$_2$-II
at around $\Ha\approx 2$, where the branch 3T-MRW$_2$-I loses
stability, seemingly by means of a saddle-node bifurcation. This is
because taking an initial condition on branch 3T-MRW$_2$-I and
integrating with $\Ha>2$, a solution on the branch 3T-MRW$_2$-II is
reached. The unstable part of 3T-MRW$_2$-I, born at the saddle node,
may extend for smaller $\Ha=1.7$, and thus the transients close to the
chaotic saddle may approach this branch and be repelled afterwards to
branch 3T-MRW$_2$-II, thanks to the connection of their invariant
manifolds. Aside from $t\in[15,20]$, the approach of the transient to
branch 3T-MRW$_2$-I and the subsequent repelling to branch
3T-MRW$_2$-II can be also identified from $t\in[40,44]$.

\begin{figure}
  \hspace{0.mm}\includegraphics[width=0.99\linewidth]{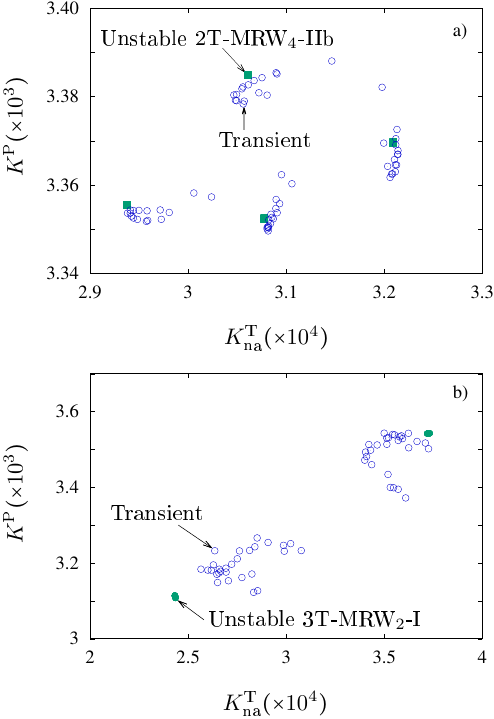}
\caption{{\bf Chaotic transient at $\Ha=1.3$:} Poincar\'e sections defined by
  the constraint $K(t)=\overline{K}$, $K$ being the volume-averaged
  kinetic energy and $\overline{K}$ its time average. The
  volume-averaged poloidal kinetic energy $K^{\text{P}}$ is displayed
  versus the volume-averaged toroidal non-axisymmetric energy
  $K^{\text{T}}_{\text{na}}$. (a) Poincar\'e sections for the
  transient (restricted to the time interval (b) of
  Fig.~\ref{fig:Ha1.3ener}, circles) and for the unstable
  2T-MRW$_4$-IIb (squares).  (b) Poincar\'e sections for the transient
  (restricted to the time interval (c) of Fig.~\ref{fig:Ha1.3ener},
  circles) and for the unstable three tori 3T-MRW$_2$-I (full circle).
  Because volume-averaged properties are considered, the Poincar\'e
  sections of 2T and 3T are a single point and a closed curve,
  respectively. The closed curve in (b) is so small that looks like a
  point, but it is not.}
\label{fig:Ha1.3poinc}
\end{figure}

\begin{figure}
  \hspace{0.mm}\includegraphics[width=0.99\linewidth]{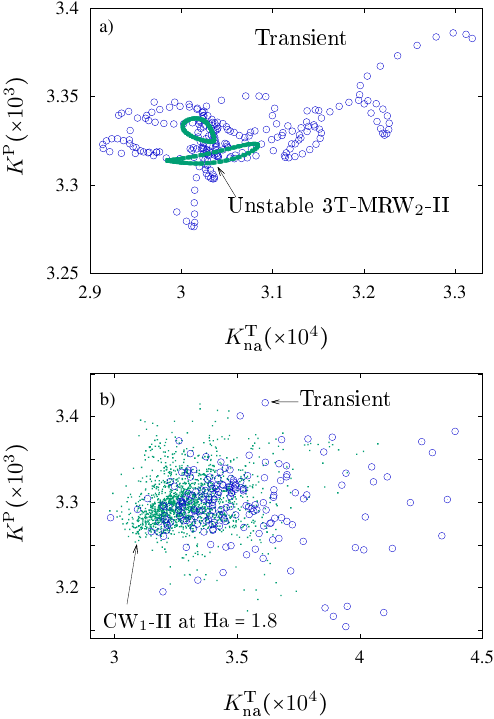}
\caption{{\bf Chaotic transient at $\Ha=1.7$:} Poincar\'e sections
  defined by the constraint $K(t)=\overline{K}$, $K$ being the
  volume-averaged kinetic energy and $\overline{K}$ its time
  average. The volume-averaged poloidal kinetic energy $K^{\text{P}}$
  is displayed versus the volume-averaged toroidal non-axisymmetric
  energy $K^{\text{T}}_{\text{na}}$. (a) Poincar\'e sections for the
  transient (restricted to the time interval (b) of
  Fig.~\ref{fig:Ha1.7ener}, circles) and for the unstable
  3T-MRW$_2$-II (full circle).  (c) Poincar\'e sections for the
  transient (restricted to the time interval (c) of
  Fig.~\ref{fig:Ha1.7ener}, circles) and for the stable CW$_1$-II
  (dots) at $\Ha=1.8$.  Because volume-averaged properties are
  considered, the Poincar\'e sections of a 3T are a closed curve.}
\label{fig:Ha1.7poinc}
\end{figure}

As in Fig.~\ref{fig:Ha0.0poinc}, the Poincar\'e sections of
volume-averaged kinetic energies are displayed in
Fig.~\ref{fig:Ha1.3poinc} and Fig.~\ref{fig:Ha1.7poinc} for the
transients found at $\Ha=1.3$ and $\Ha=1.7$, respectively,
corresponding to the chaotic saddle II. The four square points of
Fig.~\ref{fig:Ha1.3poinc}(a) correspond to the unstable 2T-MRW$_4$-IIb
with period doubled and are surrounded by four patches of circle
points corresponding to the transient during the time interval
labelled with (b) in Fig.~\ref{fig:Ha1.3ener}(b). Analogously, the
circles corresponding to the transient sections, during the time
interval labelled with (c) in Fig.~\ref{fig:Ha1.3ener}(c), are located
very close to the two closed curves (which can be barely
distinguished) of Fig.~\ref{fig:Ha1.3poinc}(b). The similarity between
the Poincar\'e sections of the transient and the unstable objects is
repeated for $\Ha=1.7$ (see Fig.~\ref{fig:Ha1.7poinc}). In this case,
the sections for the unstable 3T-MRW$_2$-II are clearly two closed
curves, lying very close (see Fig.~\ref{fig:Ha1.7poinc}(a)). We recall
(as commented in Sec.~\ref{sec:brII}) that by increasing $\Ha$ along
branch 3T-MRW$_2$-II the two closed curves overlap, giving rise to a
tangency of the orbit in the phase space. The overlapping of the two
patches of the Poincar\'e section is large for the chaotic attractor
of CW$_1$-II at $\Ha=1.8$ since these chaotic flows bifurcate from the
branch 4T-MRW$_1$-II, for which the overlapping of the section already
occurs. Accordingly, the overlapping of the transient at $\Ha=1.7$ is
also large (see Fig.~\ref{fig:Ha1.7poinc}(b)).

The overlapping of the Poincar\'e sections may indicate a riddled
basin of attraction (\cite{SoOt93,He05}), which may favour
intermittent behaviour for transients corresponding to chaotic saddles
emanating from branch II. This overlapping is not present on branch I
(e.\,g. see Fig.~\ref{fig:3d_poinc}(a,c)), thereby perhaps inhibiting
intermittency during the transients of chaotic saddles I. As shown in
the next section, the intermittent nature of the transients
corresponding to branch II is confirmed.

\section{On-off intermittency statistics}
\label{sec:int_stat}

The previous section has provided evidence that the transients within
the chaotic saddle II exhibit intermittent behaviour. During some time
intervals, these transients have values of kinetic energy of the modes
$m\ne 4k$ significantly smaller than those corresponding to the
kinetic energy of the modes $m=4k$, for some integer $k$ (see
Fig.~\ref{fig:mener}(b)). This basically means that during these time
intervals, the transient almost lies within the invariant manifold
defined by the azimuthal symmetry $m=4$ and is derailed from this
manifold when the kinetic energies of the mode $m=1,2,3$ are
significant.  This type of intermittent behaviour, in which the orbit
is subsequently directed and repelled from certain unstable attractors
lying within a invariant manifold (also quasi-invariant), has been
defined by~\cite{PST93} as on-off intermittency. This type of
intermittent chaos has been widely characterised
(e.\,g.~\cite{HPH94,PHH94,APM05}) and their statistical signature can
be extracted from time series (e.\,g.~\cite{VAOS95,VAOS96,TPS02}).

\begin{figure}
  \hspace{0.mm}\includegraphics[width=0.99\linewidth]{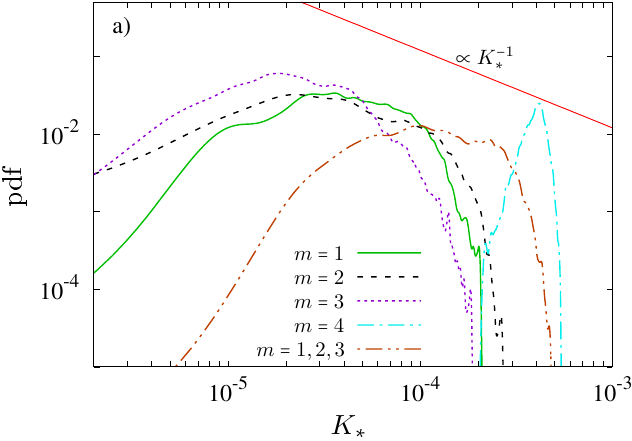}
\caption{Probability distribution function of the kinetic energy $K_*$
  of the individual modes $m=1,2,3,4$ for a intermittent transient
  flow at $\Ha=0.9$. For $m=1,2,3$ there is a range of $K_*$ for which
  the pdf has slope $\gamma \in(-1,0)$. This is not the case for
  $m=4$. The theoretical scaling is marked with a solid line (red
  online).}
\label{fig:stat_Ha0.9}
\end{figure}

In this section, some statistical analyses are performed on the time
series of transients close to either the chaotic saddle I (at
$\Ha=0.0$) or the chaotic saddle II (at $\Ha=0.9,1.3,1.7$), but also
to the time series of a stable chaotic attractor on branch CW$_1$-II
(at $\Ha=1.8$). The analyses are done for $K_*=K_1+K_2+K_3$ as a
measure of the departure of the intermittent transient from the
invariant subspace of azimuthal symmetry $m=4$. Similar results are
obtained when $K_*=K_i$ and $i=1,2,3$. For on-off intermittency, the
probability measuring the departure from the invariant manifold should
have a power law distribution $P(K_*)\sim K_*^{\gamma}$ with $\gamma
\in(-1,0)$ (\cite{VAOS96,TPS02}).  Figure~\ref{fig:stat_Ha0.9}
displays the probability distribution function (pdf) of the kinetic
energy $K_*$ of the modes $m=1,2,3,4$ and the sum of the kinetic
energies of the modes $m=1,2,3$ ($K_*=K_1+K_2+K_3$) for the transient
at $\Ha=0.9$.  Either for $K_*=K_i$, $i=1,2,3$, or for
$K_*=K_1+K_2+K_3$, there is an interval of $K_*$ values for which the
pdf has a slope $\gamma \in(-1,0)$, which is a signature of on-off
intermittency for the transient at $\Ha=0.9$. This is clearly not the
case for $m=4$ since $K_4$ does not measure the departure from the
invariant manifold, which is defined by the $m=4$ azimuthal symmetry.

\begin{figure*}
  \hspace{0.mm}\includegraphics[width=0.99\linewidth]{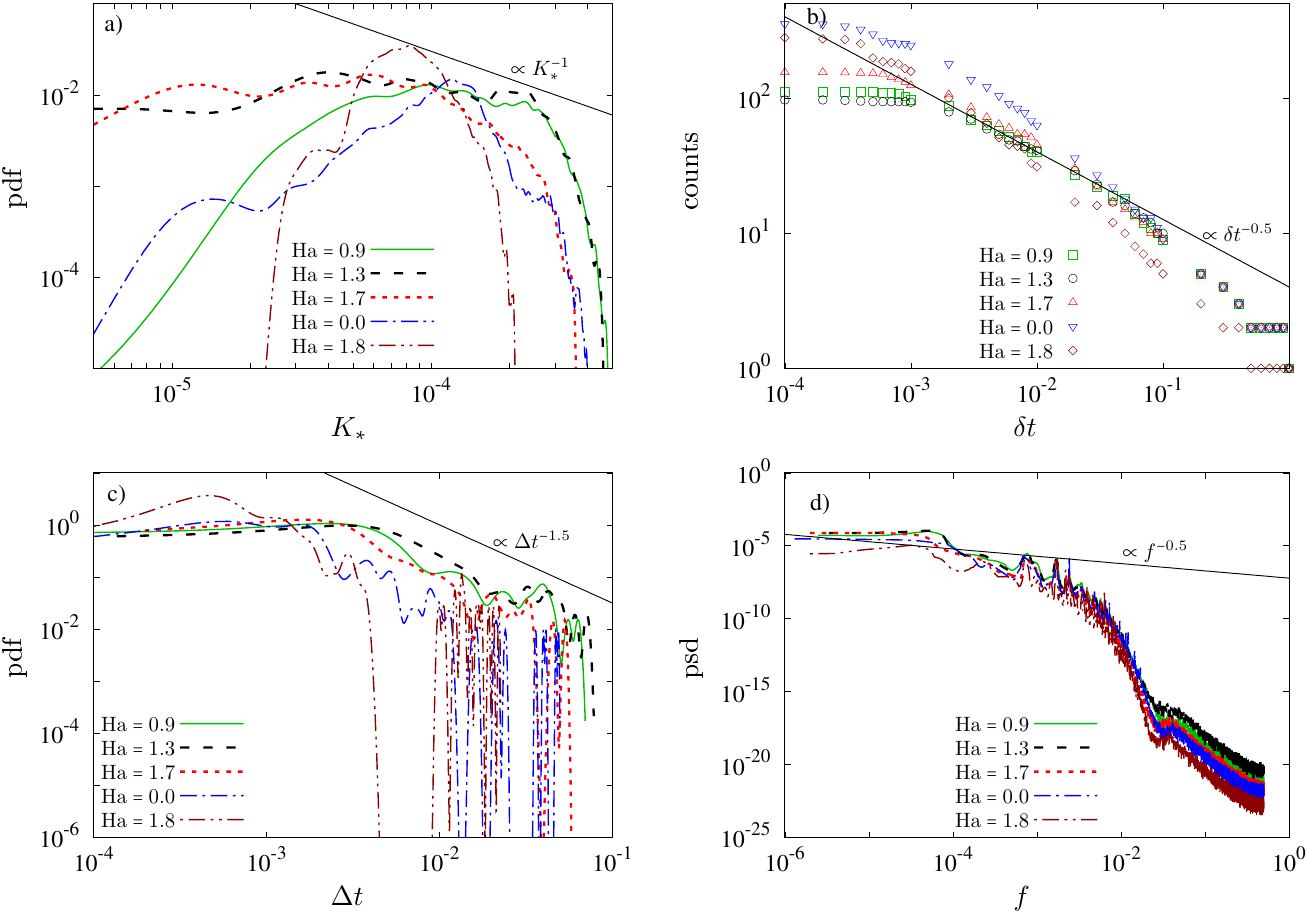}
\caption{(a) Probability distribution function of the sum of the
  kinetic energies $K_*$ contained in the modes $m=1,2,3$ for
  intermittent transient flows. For $\Ha=0.9,1.3,1.7$ there is a range
  of $K_*$ for which the pdf has slope $\gamma \in(-1,0)$. This is not
  the case for $\Ha=0$ or $\Ha=1.8$. (b) Number of counts versus
  length of the interval for the set of burst times. (c) Probability
  distribution function of the interburst times (d) Power spectral
  density of $K_*$. The theoretical scalings are marked with a solid
  straight line.}
\label{fig:stat_m1-3}
\end{figure*}

Figure~\ref{fig:stat_m1-3}(a) displays the pdf of $K_*=K_1+K_2+K_3$
for all the time series considered: a transient within the chaotic
saddle I ($\Ha=0.0$), transients close to the chaotic saddle II
($\Ha=0.9,1.3,1.7$), and a stable solution from branch CW$_1$-II
($\Ha=1.8$). While for $\Ha=0.9,1.3,1.7$ (i.\,e. transients within the
chaotic saddle II) the pdf seems to follow the predicted power laws
for on-off intermittency, these can not be identified in the pdf of
either the transient close to the chaotic saddle I or the stable
CW$_1$-II, suggesting that these two cases do not correspond to on-off
intermittency. This seems also to be confirmed with other on-off
intermittency statistical indicators explained as follows.

With time series analysis, a threshold $K_c=\alpha \max{K_*}$, with
$\alpha\sim O(0.1)$, is considered to distinguish between the on and
off phases. This threshold defines the set of burst times as the times
when the curve of $K_*$ crosses the threshold line in the upper
direction so the solution is away from the $m=4$ invariant subspace
(see \cite{VAOS96}). The factors $\alpha$ for $\Ha=0$, $\Ha=0.9$,
$\Ha=1.3$, $\Ha=1.7$, and $\Ha=1.8$, are $\alpha=0.25$, $\alpha=0.26$,
$\alpha=0.4$, $\alpha=0.28$, and $\alpha=0.34$, respectively, and the
maximum values are $\max{K_*}=4\times 10^{-4}$, $\max{K_*}=4.7\times
10^{-4}$, $\max{K_*}=4.51\times 10^{-4}$, $\max{K_*}=3.56\times
10^{-4}$, and $\max{K_*}=2.08\times 10^{-4}$, respectively. The set of
burst times (normalised) has a fractal box-counting dimension $d=1/2$
(\cite{VAOS96}), i.\,e. the number $N$ of time intervals of length
$\delta t$ required to cover the fractal set follows $N(\delta t)\sim
\delta t^{-1/2}$. This is displayed in Figure~\ref{fig:stat_m1-3}(b)
and the agreement with the on-off intermittency theoretical scaling is
again only valid for $\Ha=0.9,1.3,1.7$ (i.\,e. for transients close to
the chaotic saddle II).

Another characteristic signature of on-off intermittency is provided
by the notion of the set of interburst times (defined as the
difference $\Delta t$ of two successive burst times). The pdf of the
interburst times follows the scaling $P(\Delta t)\sim \Delta t^{-3/2}$
(\cite{VAOS96}). Figure~\ref{fig:stat_m1-3}(c) displays these pdf for
the analysed time series. For $\Ha=0.9,1.3,1.7$ (especially for
$\Ha=1.3$) the pdf seem to approach the valid scaling for on-off
intermittent behaviour. Notice that the agreement with the theory is
not clear as in Fig.~\ref{fig:stat_m1-3}(b) since very large time
series are required to approximate the set of interburst times and the
time series are limited by the lifetime of the transients. The same
shortcoming occurs when considering the power spectral density (psd)
of the time series of $K_*$. In the case of on-off intermittent signal,
the psd scales as $f^{-1/2}$ (\cite{VAOS95}), which can not be clearly
identified from our time series (see figure~\ref{fig:stat_m1-3}(d)).

\section{Conclusions}
\label{sec:con}

The study focuses on the analysis of bifurcation phenomena between
regular and chaotic magnetohydrodynamic (MHD) flows. These flows are
obtained by direct numerical simulations of the magnetised spherical
Couette (MSC) system, a widely used three-dimensional MHD model for
the study of astrophysical phenomena. As recently found
by~\cite{GSGS20b}, symmetry breaking Hopf bifurcations may give rise
to stable attractors described by four fundamental frequencies
(i.\,e. four-dimensional invariant tori). According to~\cite{NRT78},
this scenario is not possible when the system is generic
(i.\,e. without any prescribed symmetry) since any perturbation
applied to three-dimensional invariant tori gives rise to chaotic
behaviour. The analysis of spatially symmetric systems, as the MSC
system, is thus of fundamental importance to understand the type of
bifurcations that may occur.

The mechanism, first described in~\cite{GSGS20b}, of the generation of
four-dimensional invariant tori (4T) in the form of modulated rotating
waves (MRW) is confirmed in the present study by analysing a new
branch of solutions. As in~\cite{GSGS20b}, the bifurcation parameter
is the Hartmann number, measuring the strength of the magnetic field
applied to the system, while the other parameters (the Reynolds number
and aspect ratio of the shell) are fixed. The results show that the
new branch (branch II) giving rise to 4T follows the same sequence of
Hopf bifurcations as the branch described in~\cite{GSGS20b} (branch
I), but in contrast, the structure of solutions in the phase space is
significantly different for both branches. By analysing Poincar\'e
sections of 3T-MRW and 4T-MRW on both branches, we have revealed the
different phase space distributions of their solutions. In the case of
branch II, the two patches of the Poincar\'e sections overlap,
indicating a tangency of the orbit (\cite{He05}), which may indicate a
riddled basin of attraction. In contrast to this, the two patches of
the Poincar\'e sections of solutions along branch I are clearly
separated and the three-dimensional phase space plots of the orbit
have a toroidal-like structure.

With the help of very long time series, we have been able to determine
the transition between 4T-MRW and chaotic flows by analysing the
diffusion of the orbit in the phase space following the procedure
described by~\cite{LFC92,Las93b}. This was not possible
in~\cite{GSGS20b} because the analysed time series were not
long enough. We have shown that before the transition, 4T-MRW
may develop resonances, i.\,e. one of the fundamental frequencies can
be expressed as linear combination of the others. This is verified by
employing an accurate algorithm (\cite{Las93}) for determining the
fundamental frequencies.

Aside from the study of high dimensional invariant tori and their
transition to chaos on branches I and II, another important goal of
this research is to investigate in detail the nature and prevalence of
chaotic flows when varying the Hartmann number. Chaotic MHD attractors
only appear to be stable in a small range of Hartmann numbers close to
the transition from regular 4T-MRW. Away from the range of stability
of chaotic attractors on both branches (I and II), the time
integrations typically involve very long initial transients (which may
exceed 50 dimensionless units) before a stable attractor (which may
not belong to branch I nor branch II) is reached. The existence of
these long initial transients has been interpreted as a result of a
crisis, at which a stable chaotic attractor loses stability developing
a chaotic saddle.

By a direct comparison of time series of volume-averaged kinetic
energies and Poincar\'e sections of the initial transients with
unstable MRW with either azimuthal symmetry $m=4$ or $m=2$, existing
at the same parameters, we have demonstrated that during the
transients, the phase space trajectory approaches these unstable MRW
at certain periods, but after some time interval it is repelled
farther away. The different phase space structure of unstable MRW
belonging to branch I and branch II make the corresponding transients
also different. Specifically, the tangency in the phase space observed
for 3T-MRW belonging to branch II favours the development of
intermittent behaviour (e.\,g. \cite{He05}), which is observed during
the transients associated with the chaotic saddle of branch II, but
not during the transients associated with branch I.

The intermittent nature of the transients is investigated by employing
well-known statistical methods, including the analysis of the set of
burst times, designed for the time series. In the case of transients
associated with chaotic saddles of branch II, the intermittency is of
on-off type whereas this is not the case for transients associated
with branch I as commented earlier. On-off intermittency involves the
existence of an invariant manifold, at which the intermittent
trajectory is attracted, with some unstable transverse direction, from
which the trajectory is subsequently repelled. In our case the on-off
invariant manifold corresponds to the azimuthal symmetry $m=4$.

On-off intermittency may arise from a blow-out bifurcation
(\cite{OtSo94}), but as commented in Sec.~\ref{sec:brII}, this seems
to not be the case. In our situation, the existence of a riddled basin
of attraction for the chaotic flows on branch II and the intricate
structure of the invariant manifolds of the unstable 3T-MRW$_2$-I and
3T-MRW$_2$-II branches for $\Ha\in[0.9,1.7]$ allow the transient orbit
to jump and approach the different branches, thus explaining the
observed on-off intermittent signature for transients corresponding to
chaotic saddle II. Conversely, the absence of this riddled basin or
connection between branches I and II may also explain why the on-off
intermittent signature is not found for transients within the chaotic
saddle I.

Blow-out bifurcations have been found to be the mechanism for on-off
intermittent dynamics in the case of MHD problems involving the
emergence of dynamo action in a three-dimensional periodic box
(\cite{SOALF01,AlPo08}) and also in spherical geometries as is the
case of \cite{RaDo13}, which considers the same problem considered
here but without the inductionless approximation. In these studies,
the invariant manifold giving rise to the on-off intermittent
behaviour corresponds to the absence of magnetic field, which is in
contrast to our study since our manifold is defined in terms of the
azimuthal symmetry of both velocity and magnetic fields. This is
natural since in our case a magnetic field is always present, although
it is very weak since the range of Hartmann numbers investigated is
order of unity.

\begin{acknowledgments}
This project has received funding from the European Research Council
(ERC) under the European Union’s Horizon 2020 research and innovation
programme (grant agreement No 787544).
\end{acknowledgments}





\bibliographystyle{elsarticle-harv}






%

\end{document}